\documentclass[a4paper,fleqn,usenatbib]{mnras}

\usepackage[T1]{fontenc}
\usepackage{ae,aecompl}

\usepackage{graphicx}	
\usepackage{amsmath}	
\usepackage{amssymb}	
\usepackage{threeparttable}
\usepackage{multirow}
\usepackage[T1]{fontenc}
\usepackage{aecompl}

\newcommand{\tabincell}[2]{\begin{tabular}{@{}#1@{}}#2\end{tabular}}

\title[Constraining External Reverse Shock Physics of GRBs from ROTSE-III Limits]{Constraining External Reverse Shock Physics of GRBs from ROTSE-III Limits}

\author[X. H. Cui et al.]{
Xiao-Hong Cui,$^{1,2}$\thanks{E-mail: xhcui@bao.ac.cn (XHC)}
Yuan-Chuan Zou,$^{3}$
Jun-Jie Wei,$^{2}$
Wei-Kang Zheng,$^{4}$
\newauthor and Xue-Feng Wu$^{2,5}$\thanks{      xfwu@pmo.ac.cn (XFW)}\\
$^{1}$Key Laboratory of Optical Astronomy, National Astronomical Observatories, Chinese Academy of Sciences,
             Beijing 100012, China\\
$^{2}$Purple Mountain Observatory, Chinese Academy of Sciences, Nanjing 210008, China\\
$^{3}$School of Physics, Huazhong University of Science and Technology, Wuhan 430074, China\\
$^{4}$Department of Astronomy, University of California, Berkeley, CA 94720-3411, USA\\
$^{5}$Joint Center for Particle, Nuclear Physics and Cosmology, Nanjing University-Purple Mountain Observatory, Nanjing 210008, China
}

\date{Accepted XXX. Received YYY; in original form ZZZ}

\pubyear{2017}

\begin{document}
\label{firstpage}
\pagerange{\pageref{firstpage}--\pageref{lastpage}}
\maketitle

\begin{abstract}
Assuming that the early optical emission is dominated by the external reverse shock (RS) in the standard model of gamma-ray bursts (GRBs), we intend to constrain RS models with the initial Lorentz factor $\Gamma_0$ of the outflows based on the ROTSE-III observations. We consider two cases of the RS behavior: the relativistic shock and the non-relativistic shock. For homogeneous interstellar medium (ISM) and wind circum-burst environment, the constraints can be achieved by the fact that the peak flux $F_{\rm \nu}$ at the RS crossing time should be lower than the observed upper limit $F_{\rm \nu,limit}$. We consider the different spectral regimes that the observed optical frequency $\nu_{\rm opt}$ may locate in, which are divided by the orders for the minimum synchrotron frequency $\nu_{\rm m}$ and the cooling frequency $\nu_{\rm c}$. Considering the homogeneous and wind environment around GRBs, we find that the relativistic RS case can be constrained by the (upper and lower) limits of $\Gamma_0$ in a large range from about hundreds to thousands for 36 GRBs reported by ROTSE-III. The constraints on the non-relativistic RS case are achieved with limits of $\Gamma_0$ for 26 bursts ranging from $\sim 30$ to $\sim 350$. The lower limits of $\Gamma_0$ achieved for the relativistic RS model is disfavored based on the previously discovered correlation between the initial Lorentz factor $\Gamma_0$ and the isotropic gamma-ray energy $E_{\rm \gamma, iso}$ released in prompt phase.
\end{abstract}




\section{Introduction}

The standard afterglow model (M{\'e}sz{\'a}ros \& Rees 1997; Sari, Piran, \& Narayan 1998) of gamma-ray bursts (GRBs) has been proved to be successful in explaining the observed broadband afterglows. According to this model, a forward shock (FS) and a reverse shock (RS) will form when the ejecta of a GRB sweeps up an interstellar medium (ISM) surrounding the GRB. The afterglows are well described by synchrotron radiation of non-thermal electrons accelerated in both shocks. With optical data collections and analyses (e.g., Liang \& Zhang 2006; Nardini et al. 2006; Kann et al. 2006; Oates et al. 2009; Li et al. 2012; Liang et al. 2013; Zaninoni et al. 2013; Wang et al. 2013), the strong optical flashes associated with GRBs, e.g. 990123 (Akerlof et al. 1999; Sari \& Piran 1999a; M{\'e}sz{\'a}ros \& Rees 1999; Kobayashi \& Sari 2000), GRB 061126 (Perley et al. 2008), and GRB 130427A (Vestrand et al. 2014; Laskar et al. 2013; Perley et al. 2014), can be attributed to the emission from an RS. In particular, the evidence of RS emission are reported in GRB 130427A (Laskar et al. 2013; Perley et al. 2014) based on not only optical but radio data predicted early by Soderberg \& Ramirez-Ruiz (2002, 2003).

The prompt $\gamma$-ray emission from GRBs is generally believed to be produced in a relativistic jet with an initial bulk Lorentz factor ($\Gamma_0$). But the origin of jet is not well understood mostly due to the lack of knowledge of its composition, energy dissipation and the mechanism of particle acceleration, e.g., Kumar \& Zhang (2015) for a review. In a widely used fireball model, a GRB could be produced in the mergers of binary neutron stars (Narayan et al. 1992) or collapses of massive stars (Woosley 1993). Although there are two types of widely discussed central engines: a hyper-accreting stellar-mass black hole (e.g. Woosley 1993; Popham et al. 1999; Wang et al. 2002; Chen \& Beloborodov 2007; Lei et al. 2009, 2013; Nagataki 2009, 2011) and a ``fast-spinning magnetar'' (e.g. Usov 1992; Thompson 1994; Dai \& Lu 1998a; Zhang \& M{\'e}sz{\'a}ros 2001; Dai et al. 2006; Bucciantini et al. 2008, 2009; Metzger et al. 2011), they have not been identified yet. It is still unclear how the jet is launched from the GRB central engine. The study about the initial Lorentz factor of the jet is important because it carries the information about the jet launching, baryon loading process, photosphere emissions and links the broadband observational data with the standard fireball model (e.g., Lei et al. 2013; Yi et al. 2017).

As a crucial parameter to understand the physics of shock and to constrain the models of GRB sources, the initial Lorentz factor $\Gamma_0$ of GRB shells during the prompt $\gamma$-ray emission phase in principle can be retrieved from the reverse-shock data (Sari \& Piran 1995). This shock heats up the shell's material and operates only once, producing a single flash of a duration comparable to the duration of the GRB (Sari \& Piran 1999b). Therefore, after the shock passes through the shell, no more emission above the cooling frequency will be produced. There are two possibilities for the RS (Sari \& Piran 1995, 1999b), one is the non-relativistic case in which the shock is non-relativistic when it begins to cross the shell, and the other one is the relativistic case in which the shock is relativistic when it crosses the shell. The difference between the two cases depends on the shock conditions. The typical synchrotron frequency of an RS is the minimum synchrotron frequency $\nu_{\rm m}$. If the frequency $\nu_{\rm m}$ and the cooling frequency $\nu_{\rm c}$ fall in the optical region, a strong optical emission can be observed and can be used to estimate the initial Lorentz factor $\Gamma_0$ (Sari \& Piran 1999a; Zhang, Kobayashi \& M{\'e}sz{\'a}ros 2003; Molinari et al. 2007; Jin \& Fan 2007; Hasco{\"e}t et al. 2015). Consequently, the RS models can be constrained by these estimations of $\Gamma_0$. If the optical emission is not observed but only a limit is reported, the model constraint can still be achieved from the limits of some of the physical parameters, e.g. the initial Lorentz factor. In this paper, we constrain the RS models in a uniform and a wind environment with the initial Lorentz factor of GRBs from ROTSE III limits.

Except for from the peak time of RS emission, a lower limit of the initial Lorentz factor $\Gamma_0$ can be deduced from the ``compactness
problem'' (e.g. Fenimore, Epstein \& Ho 1993; Woods \& Loeb 1995; Baring \& Harding 1997; Lithwick \& Sari 2001; Abdo et al. 2009; Ackermann et al. 2010; Aoi et al. 2010; Li 2010; Zhao et al. 2011; Zou et al. 2011; Tang et al. 2015; Hasco{\"e}t et al. 2015). If the peak time of the afterglow lightcurve corresponds to the decelerating time of the ejecta (M{\'e}sz{\'a}ros \& Rees 1997; Sari \& Piran 1999a,b; Kobayashi 2000), the value of $\Gamma_0$ can be estimated based on the observations of this peak at the optical bands (e.g. Rykoff et al. 2009) or in the X-ray band (e.g. Xue, Fan \& Wei 2009). Within the ``internal--external'' shock scenario, the brightness of the early FS emission depends sensitively on
the initial Lorentz factor. Using this model, Zou \& Piran (2010) set upper limits of $\Gamma_0$ for the variant bursts. In addition, the correlation between the initial Lorentz factor $\Gamma_0$ and the burst isotropic energy $E_{\rm \gamma, iso}/$luminosity $L_{\gamma}$ was found based on the early afterglow observations of GRBs (Liang et al. 2010; L\"u et al. 2012; Hasco{\"e}t et al. 2014). Using the relation of $\Gamma_0-L_{\gamma}$, Mu et al. (2016) estimated the Lorentz factors of 29 bright X-ray flares. The Lorentz factor of the late X-ray flares can also be determined by the thermal emissions of the flares (Peng et al. 2014) or/and curvature effect in the decay phases (Jin et al. 2010; Yi et al. 2015). Moreover, Sonbas et al. (2015) reported a mutual correlation among the minimum variability timescale, the spectral lag of the prompt emission and the bulk initial Lorentz factor.

In this paper, we adopt the early optical data obtained by ROTSE-III\footnote{http://www.rotse.net/} (Akerlof et al. 2003), which consists of four 0.45-m robotic reflecting telescopes with a field of view (FoV) of $1.85^{\circ}\times 1.85^{\circ}$ for each and is managed by a fully-automated system.  With large FoV and fast slewing abilities, small robotic telescopes like ROTSE-III are able to detect the prompt optical emission in minutes time scale after the trigger of a GRB by $\gamma$-ray detectors. In the next section, we will present our method and ROTSE-III data of analysis. The relativistic-RS and nonrelativistic-RS cases in homogeneous ISM and wind environment are used to deduce the constraints of the RS models with the initial Lorentz factor $\Gamma_0$ considering the effect of redshift. In \S~3, we will present the results of constraints from the upper and lower limits of the initial Lorentz factor $\Gamma_0$ for the ROTSE-III sample. Conclusions and discussion will be presented in \S~4.

\section{Methodology and Observational data}

The interactions of a relativistic fireball with surrounding matter can be described by two shocks: an RS that propagates back into the ejecta of the fireball and an FS that propagates into the ambient medium. Our model is based on basic consideration of this RS-FS system. The emission of very early afterglow from the RS in homogeneous ISM and wind environment are studied here. From optical observation of ROTSE-III limits, we constrain the RS models with the initial Lorentz factor $\Gamma_0$. Considering the observed frequency $\nu_{\rm opt}$ in the different spectral regimes by the minimum synchrotron frequency $\nu_{\rm m}$ and the cooling frequency $\nu_{\rm c}$, the model constraints with $\Gamma_0$ are determined by the condition that the peak flux $F_{\nu}$ at the shock crossing time should be less than the observed upper limit of flux $F_{\rm \nu,limit}$, i.e. $F_{\rm \nu} < F_{\rm \nu,limit}$.

\subsection{Models}
From the study of Sari \& Piran (1995), if the ratio of the GRB shell density to the density of circum-burst medium is high, the RS will be a non-relativistic RS (NRS); and if the ratio is low, the RS will be a relativistic RS (RRS) and it considerably decelerates the shell material. Considering a particle number density $n_0$ of ISM, a shell with an isotropic kinetic energy  \emph{E}, an initial Lorentz factor $\Gamma_0$ and a width in laboratory frame $\Delta_0$ is ejected from the explosion center into ISM. The fractions of the shock energy going into the electrons and magnetic field are described by the parameters $\epsilon_{\rm e}$ and $\epsilon_{\rm B}$ respectively. Applying the Sedov length $l$, Sari \& Piran (1995) showed that the RS becomes relativistic and begins to decelerate the shell material if the shell is thick ($\Delta_0>l/2\Gamma_0^{8/3}$). The RS can be non-relativistic initially if the ISM density is low enough or the shell is thin ($\Delta_0<l/2\Gamma_0^{8/3}$). The non-relativistic RS will develop to be mildly relativistic lately. The standard radiation mechanism for GRB afterglows is synchrotron radiation by relativistic electrons accelerated by the shock into a power-law energy distribution with an index $p$ in the magnetic field, which has been studied by Kobayashi (2000). Based on the work of Kobayashi (2000) and combining with the observations of ROTSE-III, here we give the constraints on the RS models with the initial Lorentz factor $\Gamma_0$ in the thick and thin shell cases considering the effect of the redshift $z$. Only from observed optical limits, we can not know the RS producing the optical emissions is the NRS or the RRS. Both cases, i.e. the RRS (thick shell) and the NRS (thin shell), are applied separately in this work. We discuss the possible cases that the observed optical frequency $\nu_{\rm opt}$ is in the different spectral regimes by the two break frequencies, i.e., the minimum synchrotron frequency $\nu_{\rm m}$ and the cooling frequency $\nu_{\rm c}$.

The early afterglows in the wind environments have been studied by several works (Dai \& Lu 1998b; M{\'e}sz{\'a}ros et al. 1998; Chevalier \& Li, 1999, 2000; Panaitescu \& Kumar 2000, 2004; Wu et al. 2003, 2004; Kobayashi \& Zhang 2003; Zou et al. 2005; Zou \& Piran 2010; Lei et al. 2011; Yi et al. 2013) and reviewed by Gao et al. (2013). In a wind environment, the number density of surrounding medium would not be a constant but decreases with the square of the radius, i.e. $n = A r^{-2}$, where $A \approx 3\times 10^{35} A_*$ cm$^{-1}$ and $A_*$ is the wind parameter. The synchrotron self-absorption effect is not considered in this work since the absorption frequency $\nu_{\rm a}$ is assumed to be smaller than the optical frequency for typical values of physical parameters.

1. the RRS case

The peak time of the emission from the RS is comparable to the GRB duration $T_{90}$. The RS crosses the GRB shell at a time about $t_{\oplus} \approx T_{90}$. The break frequencies and the peak flux at the peak time for a homogeneous ISM can be estimated by (Sari \& Piran 1999a, b; Kobayashi 2000),
\begin{equation}
\label{numA}
\nu_{\rm m}=9.2\times 10^{12}(1+z)^{-1}\Gamma_{0,2}^2\bar{\epsilon}_{\rm e,-1}^2\epsilon_{\rm B,-2}^{1/2}n_0^{1/2}{\rm Hz},
\end{equation}
\begin{equation}
\label{nucA}
\nu_{\rm c}=3.3\times 10^{17}(1+z)^{-1/2} E_{52}^{-1/2}t_{\oplus}^{-1/2}\epsilon_{\rm B,-2}^{-3/2}(1+Y)^{-2}n_0^{-1}{\rm Hz},
\end{equation}
\begin{equation}
\label{FnumA}
F_{\rm \nu,max}=11.4(1+z)^{7/4}\Gamma_{0,2}^{-1}D_{\rm L,28}^{-2}E_{52}^{5/4}t_{\oplus}^{-3/4}\epsilon_{\rm B,-2}^{1/2}n_0^{1/4}{\rm Jy},
\end{equation}
where $E_{52}=E/(10^{52}$erg), $\Gamma_{0,2}=\Gamma_0/100$, $n_0=n/(1$cm$^{-3}$), $\bar{\epsilon}_{\rm e,-1}=\epsilon_{\rm e,-1}\frac{p-2}{p-1}$, $\epsilon_{\rm e,-1}=\epsilon_{\rm e}/0.1$, $\epsilon_{\rm B,-2}=\epsilon_{\rm B}/10^{-2}$. $z$ is the redshift of GRB. $Y$ is the Compton parameter defined by the ratio of the rate of inverse Compton energy loss to the rate of synchrotron energy loss (Sari et al. 1996). $D_{\rm L,28}=D_{\rm L}/(10^{28}$ cm) and the luminosity distance $D_{\rm L}$ is calculated by adopting the cosmological parameters as $\Omega_{\rm m}=0.3$, $\Omega_\Lambda=0.7$, and $H_0=70$ km s$^{-1}$ Mpc$^{-1}$ in the following calculations.
The kinetic energy of the shell $E$ is estimated by
\begin{equation}
\label{Ek}
E=\frac{1-\eta_\gamma}{\eta_\gamma}E_{\rm \gamma,iso} \approx \frac{1-\eta_\gamma}{\eta_\gamma}\frac{4\pi D^2_{\rm L} f_\gamma}{1+z},
\end{equation}
where $\eta_{\rm \gamma}$ is the efficiency fraction of initial energy converted into observed $\gamma$-ray emissions (Lloyd-Ronning \& Zhang 2004), $E_{\rm \gamma,iso}$ is the isotropic energy of prompt $\gamma$-ray emission, and $f_{\gamma}$ is the observed $\gamma$-ray fluence. The conventional notation $Q=10^x \times Q_x$ is used throughout this paper except some special description.

The break frequencies and the peak flux for an RRS in a wind environment can be estimated with (Wu et al. 2003, Zou et al. 2005),
\begin{equation}
\label{numAW}
\nu_{\rm m}=2.6\times 10^{15}(1+z)^{-1/2}\Gamma_{0,2}^2 E_{52}^{-1/2}t_{\oplus}^{-1/2}\bar{\epsilon}_{\rm e,-1}^2 \epsilon_{\rm B,-2}^{1/2}A_{*,-1}{\rm Hz},
\end{equation}
\begin{equation}
\label{nucAW}
\nu_{\rm c}=3.0\times 10^{12}(1+z)^{-3/2} E_{52}^{1/2}t_{\oplus}^{-1/2} \epsilon_{\rm B,-2}^{-3/2} (1+Y)^{-2}A_{*,-1}^{-2}{\rm Hz},
\end{equation}
\begin{equation}
\label{FnumAW}
F_{\rm \nu,max}=2.3 \times 10^2 (1+z)^2 \Gamma_{0,2}^{-1}D_{\rm L,28}^{-2}E_{52}t_{\oplus}^{-1}\epsilon_{\rm B,-2}^{1/2}A_{*,-1}^{1/2} {\rm Jy},
\end{equation}
where $A_{*,-1}=A_*/0.1$.

Based on Equations (\ref{numA}) to (\ref{FnumAW}) and considering the observed optical frequency $\nu_{\rm opt}$ in different spectral regimes by $\nu_{\rm m}$ and $\nu_{\rm c}$, the constraints on the initial Lorentz factor of a thick shell are shown in Table \ref{Thick} for homogeneous ISM and in Table \ref{ThickW} for wind environment. The unit of the observed frequency $\nu_{\rm opt}$ is taken as the center frequency of $R$ band in standard photometric system ($\nu_{\rm R} \approx 4.3 \times 10^{14}$ Hz, Allen 1973). We define $\nu_{\rm opt, R}={\nu_{\rm opt} \over \nu_{\rm R}}$ and $f_{\gamma,-7}^*=\frac{1-\eta_\gamma}{\eta_\gamma}f_{\gamma,-7}$ in our calculations. The constraints on $\Gamma_0$ from the inequalities of ($\nu_{\rm opt}$, $\nu_{\rm c}$), ($\nu_{\rm opt}$, $\nu_{\rm m}$), and ($\nu_{\rm m}$, $\nu_{\rm c}$) are shown in the second, third and fourth columns of the tables. The constraints from the condition that upper limit of flux $F_{\rm \nu,limit}$ larger than calculated flux $F_{\rm \nu}$ from the RS model in the thick shell case, i.e. $F_{\rm \nu,limit}>F_{\rm \nu}$, are presented in the last column of both tables. When $\nu_{\rm opt}$ in different spectral regimes by $\nu_{\rm m}$ and $\nu_{\rm c}$, the constraint on the initial Lorentz factor will be achieved from the common region where all three inequalities presented in each row of the table are satisfied. For example, when $\nu_{\rm opt}<\nu_{\rm c}<\nu_{\rm m}$ as presented in the second row of Table \ref{Thick}, the constraint on $\Gamma_0$ can be deduced from the inequalities $\nu_{\rm opt} < \nu_{\rm c}$ (the second column of this row), $\nu_{\rm m}>\nu_{\rm c}$ (the fourth column of this row), and $F_{\nu}<F_{\rm \nu, limit}$ (last column of the row).
\begin{table*}
\centering
\caption{The constraints on the number density $n_0$ and the initial Lorentz factor $\Gamma_0$ in the RRS case for a homogeneous ISM. The parameter constraints from the inequalities between two frequencies, i.e., ($\nu_{\rm opt}$, $\nu_{\rm c}$), ($\nu_{\rm opt}$, $\nu_{\rm m}$), and ($\nu_{\rm m}$, $\nu_{\rm c}$), are shown in the second, third and fourth columns. The constraint on $\Gamma_0$ by the condition that the observed upper limit of flux $F_{\rm \nu,limit}$ should be higher than the theoretical flux $F_{\rm \nu}$ from the RS emission, i.e. $F_{\rm \nu,limit}>F_{\rm \nu}$, is formulated in the last column.}
\label{Thick}
\resizebox{\textwidth}{!}{ %
 \begin{tabular}{|c|c|c|c|c|}
   \hline &
 ($\nu_{\rm opt}$, $\nu_{\rm c}$) & ($\nu_{\rm opt}$, $\nu_{\rm m}$) & ($\nu_{\rm m}$, $\nu_{\rm c}$) &$F_{\nu} < F_{\rm \nu,limit}$ \\
   \hline
 $1. \nu_{\rm opt}<\nu_{\rm c}<\nu_{\rm m}$ & \tabincell{c}{$n_0 <6.9 \times 10^3 \nu_{\rm opt,R}^{-1}\epsilon_{\rm B,-2}^{-\frac{3}{2}}$\\ $(1+Y)^{-2}f_{\gamma,-7}^{*-\frac{1}{2}}D_{\rm L,28}^{-1}T_{90}^{-\frac{1}{2}}$} &--& \tabincell{c}{$\Gamma_{0,2}>5.7 \times 10^2 n_0^{-\frac{3}{4}} \bar{\epsilon}_{\rm e,-1}^{-1}\epsilon_{\rm B,-2}^{-1}$\\ $ (1+Y)^{-1}(1+z)^{\frac{1}{2}}f_{\gamma,-7}^{*-\frac{1}{4}}D_{\rm L,28}^{-\frac{1}{2}}T_{90}^{-\frac{1}{4}}$} &\tabincell{c}{$\Gamma_{0,2}>2.5 \nu_{\rm opt,R}^{\frac{1}{3}}n_0^{\frac{7}{12}}\epsilon_{\rm B,-2}(1+Y)^{\frac{2}{3}}$\\ $ (1+z)^{\frac{1}{2}} f_{\gamma,-7}^{*\frac{17}{12}}D_{\rm L,28}^{\frac{5}{6}}T_{90}^{-\frac{7}{12}}F_{\rm \nu,limit,-3}^{-1}$}\\
 \hline
 $2. \nu_{\rm c}<\nu_{\rm opt}<\nu_{\rm m}$& \tabincell{c}{$n_0 >6.9 \times 10^3 \nu_{\rm opt,R}^{-1}\epsilon_{\rm B,-2}^{-\frac{3}{2}}$\\ $(1+Y)^{-2}f_{\gamma,-7}^{*-\frac{1}{2}}D_{\rm L,28}^{-1}T_{90}^{-\frac{1}{2}}$} & \tabincell{c}{$\Gamma_{0,2}>6.9\nu_{\rm opt,R}^{\frac{1}{2}}n_0^{-\frac{1}{4}}$\\ $ \bar{\epsilon}_{\rm e,-1}^{-1}\epsilon_{\rm B,-2}^{-\frac{1}{4}}(1+z)^{\frac{1}{2}}$}&-- &\tabincell{c}{$\Gamma_{0,2}>39.8 \nu_{\rm opt,R}^{-\frac{1}{2}}n_0^{-\frac{1}{4}}\epsilon_{\rm B,-2}^{-\frac{1}{4}}(1+Y)^{-1}$\\ $(1+z)^{\frac{1}{2}}f_{\gamma,-7}^*T_{90}^{-1}F_{\rm \nu,limit,-3}^{-1}$}\\
   \hline
 $3.  \nu_{\rm c}<\nu_{\rm m}<\nu_{\rm opt}$	&	--	&	 \tabincell{c}{$\Gamma_{0,2}<6.9\nu_{\rm opt,R}^{\frac{1}{2}}n_0^{-\frac{1}{4}}$\\ $ \bar{\epsilon}_{\rm e,-1}^{-1}\epsilon_{\rm B,-2}^{-\frac{1}{4}}(1+z)^{\frac{1}{2}}$}	 &	 \tabincell{c}{$\Gamma_{0,2}>5.7 \times 10^2 n_0^{-\frac{3}{4}} \bar{\epsilon}_{\rm e,-1}^{-1}\epsilon_{\rm B,-2}^{-1}$\\ $ (1+Y)^{-1}(1+z)^{\frac{1}{2}}f_{\gamma,-7}^{*-\frac{1}{4}}D_{\rm L,28}^{-\frac{1}{2}}T_{90}^{-\frac{1}{4}}$}& \tabincell{c}{$\Gamma_{0,2}<1.7 \times 10^{-5}\nu_{\rm opt,R}^{\frac{5}{2}}n_0^{-\frac{1}{4}}\bar{\epsilon}_{\rm e,-1}^{-3}\epsilon_{\rm B,-2}^{-\frac{1}{4}}$ \\ $(1+Y)^2(1+z)^{\frac{1}{2}}f_{\gamma,-7}^{*-2}T_{90}^{2}F_{\rm \nu,limit,-3}^{2}$}\\
 \hline
 $4.\nu_{\rm opt}<\nu_{\rm m}<\nu_{\rm c}$ & --&\tabincell{c}{$\Gamma_{0,2}>6.9\nu_{\rm opt,R}^{\frac{1}{2}}n_0^{-\frac{1}{4}}$\\ $ \bar{\epsilon}_{\rm e,-1}^{-1}\epsilon_{\rm B,-2}^{-\frac{1}{4}}(1+z)^{\frac{1}{2}}$} &\tabincell{c}{$\Gamma_{0,2}<5.7 \times 10^2 n_0^{-\frac{3}{4}} \bar{\epsilon}_{\rm e,-1}^{-1}\epsilon_{\rm B,-2}^{-1}$\\ $ (1+Y)^{-1}(1+z)^{\frac{1}{2}}f_{\gamma,-7}^{*-\frac{1}{4}}D_{\rm L,28}^{-\frac{1}{2}}T_{90}^{-\frac{1}{4}}$}&\tabincell{c}{$\Gamma_{0,2}>34.9 \nu_{\rm opt,R}^{\frac{1}{5}}n_0^{\frac{1}{20}} \bar{\epsilon}_{\rm e,-1}^{-\frac{2}{5}}\epsilon_{\rm B,-2}^{\frac{1}{5}}$\\ $(1+z)^{\frac{1}{2}}f_{\gamma,-7}^{*\frac{3}{4}}D_{\rm L,28}^{\frac{3}{10}}T_{90}^{-\frac{9}{20}}F_{\rm \nu,limit,-3}^{-\frac{3}{5}}$}\\
   \hline
 $5. \nu_{\rm m}<\nu_{\rm opt}<\nu_{\rm c}$ & \tabincell{c}{$n_0 <6.9 \times 10^3 \nu_{\rm opt,R}^{-1}\epsilon_{\rm B,-2}^{-\frac{3}{2}}$\\ $(1+Y)^{-2}f_{\gamma,-7}^{*-\frac{1}{2}}D_{\rm L,28}^{-1}T_{90}^{-\frac{1}{2}}$}& \tabincell{c}{$\Gamma_{0,2}<6.9\nu_{\rm opt,R}^{\frac{1}{2}}n_0^{-\frac{1}{4}}$\\ $ \bar{\epsilon}_{\rm e,-1}^{-1}\epsilon_{\rm B,-2}^{-\frac{1}{4}}(1+z)^{\frac{1}{2}}$}	& --& \tabincell{c}{$\Gamma_{0,2}<4.4\nu_{\rm opt,R}^{\frac{3}{2}}n_0^{-\frac{5}{4}}\bar{\epsilon}_{\rm e,-1}^{-3}\epsilon_{\rm B,-2}^{-\frac{7}{4}}$\\ $(1+z)^{\frac{1}{2}}f_{\gamma,-7}^{*-\frac{5}{2}}D_{\rm L,28}^{-1}T_{90}^{\frac{3}{2}}F_{\rm \nu,limit,-3}^{2}$}\\
 \hline
 $6. \nu_{\rm m}<\nu_{\rm c}< \nu_{\rm opt}$& \tabincell{c}{$n_0 >6.9 \times 10^3 \nu_{\rm opt,R}^{-1}\epsilon_{\rm B,-2}^{-\frac{3}{2}}$\\ $(1+Y)^{-2}f_{\gamma,-7}^{*-\frac{1}{2}}D_{\rm L,28}^{-1}T_{90}^{-\frac{1}{2}}$}&--&\tabincell{c}{$\Gamma_{0,2}<5.7 \times 10^2 n_0^{-\frac{3}{4}} \bar{\epsilon}_{\rm e,-1}^{-1}\epsilon_{\rm B,-2}^{-1}$\\ $ (1+Y)^{-1}(1+z)^{\frac{1}{2}}f_{\gamma,-7}^{*-\frac{1}{4}}D_{\rm L,28}^{-\frac{1}{2}}T_{90}^{-\frac{1}{4}}$}& \tabincell{c}{$\Gamma_{0,2}<1.7 \times 10^{-5}\nu_{\rm opt,R}^{\frac{5}{2}}n_0^{-\frac{1}{4}}\bar{\epsilon}_{\rm e,-1}^{-3}\epsilon_{\rm B,-2}^{-\frac{1}{4}}$ \\ $(1+Y)^2(1+z)^{\frac{1}{2}}f_{\gamma,-7}^{*-2}T_{90}^{2}F_{\rm \nu,limit,-3}^{2}$}\\
   \hline
  \end{tabular}}
\end{table*}
%
\begin{table*}
\centering
\caption{The constraints on the wind density parameter $A_*$ and the initial Lorentz factor $\Gamma_0$ in the RRS case for a wind environment.}
\label{ThickW}
\resizebox{\textwidth}{!}{ %
 \begin{tabular}{|c|c|c|c|c|}
 \hline
 &  ($\nu_{\rm opt}$, $\nu_{\rm c}$) & ($\nu_{\rm opt}$, $\nu_{\rm m}$) & ($\nu_{\rm m}$, $\nu_{\rm c}$) &$F_{\nu} < F_{\rm \nu,limit}$ \\
 \hline
 $1.\nu_{\rm opt}<\nu_{\rm c}<\nu_{\rm m}$	&\tabincell{c}{$A_{\ast,-1}<2.8 \times 10^{-2}\nu_{\rm opt,R}^{-\frac{1}{2}}\epsilon_{\rm B,-2}^{-\frac{3}{4}} $\\ $(1+Y)^{-1}(1+z)^{-1}f_{\gamma,-7}^{*\frac{1}{4}}D_{\rm L,28}^{\frac{1}{2}} T_{90}^{\frac{1}{4}}$}	 &--	 &	 \tabincell{c}{$\Gamma_{0,2} >3.8 \times 10^{-3}A_{\ast,-1}^{-\frac{3}{2}}\bar{\epsilon}_{\rm e,-1}^{-1}\epsilon_{\rm B,-2}^{-1}$\\ $(1+Y)^{-1}(1+z)^{-1} f_{\gamma,-7}^{*\frac{1}{2}}D_{\rm L,28}T_{90}^{\frac{1}{2}}$}	& \tabincell{c}{$\Gamma_{0,2}>3.1 \times 10^{4}\nu_{\rm opt,R}^{\frac{1}{3}}A_{\ast,-1}^{\frac{7}{6}}$\\ $\epsilon_{\rm B,-2}(1+Y)^{\frac{2}{3}}(1+z)^{\frac{5}{3}} $\\ $f_{\gamma,-7}^{*-\frac{5}{6}}D_{\rm L,28}^{-\frac{1}{3}}T_{90}^{-\frac{7}{6}}F_{\rm \nu,limit,-3}^{-1}$}\\
   \hline
 $2.\nu_{\rm c}<\nu_{\rm opt}<\nu_{\rm m}$& \tabincell{c}{$A_{\ast,-1}>2.8 \times 10^{-2}\nu_{\rm opt,R}^{-\frac{1}{2}}\epsilon_{\rm B,-2}^{-\frac{3}{4}} $\\ $(1+Y)^{-1}(1+z)^{-1}f_{\gamma,-7}^{*\frac{1}{4}}D_{\rm L,28}^{\frac{1}{2}} T_{90}^{\frac{1}{4}}$} & \tabincell{c}{$\Gamma_{0,2}>0.1\nu_{\rm opt,R}^{\frac{1}{2}} A_{\ast,-1}^{-\frac{1}{2}}$\\ $ \bar{\epsilon}_{\rm e,-1}^{-1}\epsilon_{\rm B,-2}^{-\frac{1}{4}}f_{\gamma,-7}^{*\frac{1}{4}}
D_{\rm L,28}^{\frac{1}{2}}T_{90}^{\frac{1}{4}}  $}&-- &\tabincell{c}{$\Gamma_{0,2}>0.8\nu_{\rm opt,R}^{-\frac{1}{2}}A_{\ast,-1}^{-\frac{1}{2}}\epsilon_{\rm B,-2}^{-\frac{1}{4}}$\\ $(1+Y)^{-1}f_{\gamma,-7}^{*\frac{5}{4}}D_{\rm L,28}^{\frac{1}{2}}T_{90}^{-\frac{3}{4}}F_{\rm \nu,limit,-3}^{-1}$}\\
   \hline
 $3.\nu_{\rm c}<\nu_{\rm m}<\nu_{\rm opt}$ & --  & \tabincell{c}{$\Gamma_{0,2}<0.1\nu_{\rm opt,R}^{\frac{1}{2}} A_{\ast,-1}^{-\frac{1}{2}}$\\ $ \bar{\epsilon}_{\rm e,-1}^{-1}\epsilon_{\rm B,-2}^{-\frac{1}{4}}f_{\gamma,-7}^{*\frac{1}{4}}
D_{\rm L,28}^{\frac{1}{2}}T_{90}^{\frac{1}{4}}  $}& \tabincell{c}{$\Gamma_{0,2} >3.8 \times 10^{-3}A_{\ast,-1}^{-\frac{3}{2}}\bar{\epsilon}_{\rm e,-1}^{-1}\epsilon_{\rm B,-2}^{-1}$\\ $(1+Y)^{-1}(1+z)^{-1} f_{\gamma,-7}^{*\frac{1}{2}}D_{\rm L,28}T_{90}^{\frac{1}{2}}$}& \tabincell{c}{$\Gamma_{0,2}<4.0\times 10^{-7}\nu_{\rm opt,R}^{\frac{5}{2}}A_{\ast,-1}^{-\frac{1}{2}}$\\ $\bar{\epsilon}_{\rm e,-1}^{-3}\epsilon_{\rm B,-2}^{-\frac{1}{4}}(1+Y)^2 f_{\gamma,-7}^{*-\frac{7}{4}}$\\ $D_{\rm L,28}^{\frac{1}{2}}T_{90}^{\frac{9}{4}}F_{\rm \nu,limit,-3}^{2}$}\\
 \hline
 $4.\nu_{\rm opt}<\nu_{\rm m}<\nu_{\rm c}$	&	--	&	 \tabincell{c}{$\Gamma_{0,2}>0.1\nu_{\rm opt,R}^{\frac{1}{2}} A_{\ast,-1}^{-\frac{1}{2}}$\\ $ \bar{\epsilon}_{\rm e,-1}^{-1}\epsilon_{\rm B,-2}^{-\frac{1}{4}}f_{\gamma,-7}^{*\frac{1}{4}}
D_{\rm L,28}^{\frac{1}{2}}T_{90}^{\frac{1}{4}}  $}&	 \tabincell{c}{$\Gamma_{0,2} <3.8 \times 10^{-3}A_{\ast,-1}^{-\frac{3}{2}}\bar{\epsilon}_{\rm e,-1}^{-1}\epsilon_{\rm B,-2}^{-1}$\\ $(1+Y)^{-1}(1+z)^{-1} f_{\gamma,-7}^{*\frac{1}{2}}D_{\rm L,28}T_{90}^{\frac{1}{2}}$} &\tabincell{c}{$\Gamma_{0,2}>53.0\nu_{\rm opt,R}^{\frac{1}{5}}A_{\ast,-1}^{\frac{1}{10}}\bar{\epsilon}_{\rm e,-1}^{-\frac{2}{5}}\epsilon_{\rm B,-2}^{\frac{1}{5}} $\\ $(1+z)^{\frac{3}{5}} f_{\gamma,-7}^{*\frac{7}{10}}D_{\rm L,28}^{\frac{1}{5}}T_{90}^{-\frac{1}{2}}F_{\rm \nu,limit,-3}^{-\frac{3}{5}}$}\\
 \hline
 $5.\nu_{\rm m}<\nu_{\rm opt}<\nu_{\rm c}$ & \tabincell{c}{$A_{\ast,-1}<2.8 \times 10^{-2}\nu_{\rm opt,R}^{-\frac{1}{2}}\epsilon_{\rm B,-2}^{-\frac{3}{4}} $\\ $(1+Y)^{-1}(1+z)^{-1}f_{\gamma,-7}^{*\frac{1}{4}}D_{\rm L,28}^{\frac{1}{2}} T_{90}^{\frac{1}{4}}$}	 &\tabincell{c}{$\Gamma_{0,2}<0.1\nu_{\rm opt,R}^{\frac{1}{2}} A_{\ast,-1}^{-\frac{1}{2}}$\\ $ \bar{\epsilon}_{\rm e,-1}^{-1}\epsilon_{\rm B,-2}^{-\frac{1}{4}}f_{\gamma,-7}^{*\frac{1}{4}}
D_{\rm L,28}^{\frac{1}{2}}T_{90}^{\frac{1}{4}}  $} &--&\tabincell{c}{$\Gamma_{0,2}<4.8 \times 10^{-10}\nu_{\rm opt,R}^{\frac{3}{2}}A_{\ast,-1}^{-\frac{5}{2}} $\\ $ \bar{\epsilon}_{\rm e,-1}^{-3}\epsilon_{\rm B,-2}^{-\frac{7}{4}}(1+z)^{-2}f_{\gamma,-7}^{*-\frac{5}{4}}$\\ $ D_{\rm L,28}^{\frac{3}{2}}T_{90}^{\frac{11}{4}}F_{\rm \nu,limit,-3}^{2}$}\\
 \hline
 $6.\nu_{\rm m}<\nu_{\rm c}<\nu_{\rm opt}$ & \tabincell{c}{$A_{\ast,-1}>2.8 \times 10^{-2}\nu_{\rm opt,R}^{-\frac{1}{2}}\epsilon_{\rm B,-2}^{-\frac{3}{4}} $\\ $(1+Y)^{-1}(1+z)^{-1}f_{\gamma,-7}^{*\frac{1}{4}}D_{\rm L,28}^{\frac{1}{2}} T_{90}^{\frac{1}{4}}$}	 &--&\tabincell{c}{$\Gamma_{0,2} <3.8 \times 10^{-3}A_{\ast,-1}^{-\frac{3}{2}}\bar{\epsilon}_{\rm e,-1}^{-1}\epsilon_{\rm B,-2}^{-1}$\\ $(1+Y)^{-1}(1+z)^{-1} f_{\gamma,-7}^{*\frac{1}{2}}D_{\rm L,28}T_{90}^{\frac{1}{2}}$} &\tabincell{c}{$\Gamma_{0,2}<4.0\times 10^{-7}\nu_{\rm opt,R}^{\frac{5}{2}}A_{\ast,-1}^{-\frac{1}{2}}$\\ $\bar{\epsilon}_{\rm e,-1}^{-3}\epsilon_{\rm B,-2}^{-\frac{1}{4}}(1+Y)^2 f_{\gamma,-7}^{*-\frac{7}{4}}$\\ $D_{\rm L,28}^{\frac{1}{2}}T_{90}^{\frac{9}{4}}F_{\rm \nu,limit,-3}^{2}$}\\
    \hline
 \end{tabular}}
\end{table*}

2. the NRS case

In the thin shell case, the crossing time $t_{\oplus}$ of the shock is assumed to be the same as the peak time of the RS emissions. For homogeneous ISM, the
break frequencies and the peak flux in observer's frame can be estimated by (Sari \& Piran 1999a, b; Kobayashi 2000),
\begin{equation}
\label{numB}
\nu_{\rm m}=2.0\times 10^{2}(1+z)^{-7}\Gamma_{0,2}^{18} E_{52}^{-2}t_{\oplus}^6\bar{\epsilon}^2_{\rm e,-1}\epsilon_{\rm B,-2}^{1/2}n_0^{5/2}{\rm Hz},
\end{equation}
\begin{equation}
\label{nucB}
\nu_{\rm c}=7.1\times 10^{19}(1+z) \Gamma_{0,2}^{-4}t_{\oplus}^{-2}\epsilon_{\rm B,-2}^{-3/2}(1+Y)^{-2} n_0^{-3/2}{\rm Hz},
\end{equation}
\begin{equation}
\label{FnumB}
F_{\rm \nu,max}=9.1\times 10^{-4} (1+z)^{-1/2}\Gamma_{0,2}^5 D_{\rm L,28}^{-2}E_{52}^{1/2}t_{\oplus}^{3/2}\epsilon_{\rm B,-2}^{1/2}n_0 {\rm Jy}.
\end{equation}
The break frequencies and the peak flux in the observer's frame for wind environment can be estimated by (Wu et al. 2003, Zou et al. 2005),
\begin{equation}
\label{numBW}
\nu_{\rm m}=4.0\times 10^{14}(1+z)^{-2}\Gamma_{0,2}^8 E_{52}^{-2}t_{\oplus}\bar{\epsilon}^{2}_{\rm e,-1}\epsilon_{\rm B,-2}^{1/2}A_{*,-1}^{5/2}{\rm Hz},
\end{equation}
\begin{equation}
\label{nucBW}
\nu_{\rm c}=3.0\times 10^{12}(1+z)^{-2}\Gamma_{0,2}^{2}t_{\oplus}\epsilon_{\rm B,-2}^{-3/2}(1+Y)^{-2}A_{*,-1}^{-3/2}{\rm Hz},
\end{equation}
\begin{equation}
\label{FnumBW}
F_{\rm \nu,max}=1.3 \times 10^2 (1+z)^{3/2}\Gamma_{0,2} D_{\rm L,28}^{-2}E_{52}^{1/2}t_{\oplus}^{-1/2}\epsilon_{\rm B,-2}^{-1/2}A_{*,-1} {\rm Jy}.
\end{equation}

Based on the above equations, the constraints on the initial Lorentz factor $\Gamma_0$ in the thin shell case are presented in Table \ref{Thin} for homogeneous ISM and in Table \ref{ThinW} for wind environment. We can achieve the constraints on $\Gamma_0$ from the common region determined by three inequalities (the second to fifth columns in both Tables). The method to derive the limits of $\Gamma_0$ is the same as that of the thick shell case as described above except that the peak time of RS emission is uncertain because only the limits of magnitude are available from observations we studied here. Therefore, a time sequence after GRB prompt phase $T_{90}$ is considered in order to find the constraints on Lorentz factor $\Gamma_0$. The most conservative estimation is adopted in the NRS case. That is to say, we select the maximum of one quantity (e.g. $\Gamma_0$) we considered in the time sequence to be the lower limit of this quantity; while the minimum of quantity in the time sequence is selected to be the upper limit.
\begin{table*}
\centering
\caption{The constraints on the initial Lorentz factor $\Gamma_0$ in the NRS case for a homogeneous ISM.}
\label{Thin}
\resizebox{\textwidth}{!}{ %
 \begin{tabular}{|c|c|c|c|c|}
   \hline &
 ($\nu_{\rm opt}$, $\nu_{\rm c}$) & ($\nu_{\rm opt}$, $\nu_{\rm m}$) & ($\nu_{\rm m}$, $\nu_{\rm c}$) &$F_{\nu} < F_{\nu,limit}$ \\
   \hline
 $1. \nu_{\rm opt}<\nu_{\rm c}<\nu_{\rm m}$ & \tabincell{c}{$\Gamma_{0,2} <20.2 \nu_{\rm opt,R}^{-\frac{1}{4}}n_0^{-\frac{3}{8}}\epsilon_{\rm B,-2}^{-\frac{3}{8}}$\\ $(1+Y)^{-\frac{1}{2}}(1+z)^{\frac{1}{4}}t_{\oplus}^{-\frac{1}{2}}$} &--& \tabincell{c}{$\Gamma_{0,2}>4.2 n_0^{-\frac{2}{11}}\bar{\epsilon}_{\rm e,-1}^{-\frac{1}{11}}\epsilon_{\rm B,-2}^{-\frac{1}{11}}(1+Y)^{-\frac{1}{11}}$\\ $(1+z)^{\frac{3}{11}}f_{\gamma,-7}^{*\frac{1}{11}}D_{\rm L,28}^{\frac{2}{11}}t_{\oplus}^{-\frac{4}{11}}$} &\tabincell{c}{$\Gamma_{0,2}< 2.7 \nu_{\rm opt,R}^{-\frac{1}{19}}n_0^{-\frac{9}{38}}\epsilon_{\rm B,-2}^{-\frac{3}{19}}$\\$(1+Y)^{-\frac{2}{19}}(1+z)^{\frac{4}{19}}f_{\gamma,-7}^{*-\frac{3}{38}}$\\$ D_{\rm L,28}^{\frac{3}{19}}t_{\oplus}^{-\frac{13}{38}}F_{\rm \nu,limit,-3}^{\frac{3}{19}}$}\\
    \hline
 $2.\nu_{\rm c}<\nu_{\rm opt}<\nu_{\rm m}$& \tabincell{c}{$\Gamma_{0,2} >20.2 \nu_{\rm opt,R}^{-\frac{1}{4}}n_0^{-\frac{3}{8}}\epsilon_{\rm B,-2}^{-\frac{3}{8}}$\\ $(1+Y)^{-\frac{1}{2}}(1+z)^{\frac{1}{4}}t_{\oplus}^{-\frac{1}{2}}$} & \tabincell{c}{$\Gamma_{0,2}> 3.0 \nu_{\rm opt,R}^{\frac{1}{18}}n_0^{-\frac{5}{36}}\bar{\epsilon}_{\rm e,-1}^{-\frac{1}{9}}$\\ $\epsilon_{\rm B,-2}^{-\frac{1}{36}}(1+z)^{\frac{5}{18}}f_{\gamma,-7}^{*\frac{1}{9}}D_{\rm L,28}^{\frac{2}{9}}t_{\oplus}^{-\frac{1}{3}}$} &-- &\tabincell{c}{$\Gamma_{0,2} <0.3 \nu_{\rm opt,R}^{\frac{1}{6}}n_0^{-\frac{1}{12}}\epsilon_{\rm B,-2}^{\frac{1}{12}}$\\ $(1+Y)^{\frac{1}{3}}(1+z)^{\frac{1}{6}}f_{\gamma,-7}^{*-\frac{1}{6}}$\\ $D_{\rm L,28}^{\frac{1}{3}}t_{\oplus}^{-\frac{1}{6}}F_{\rm \nu,limit,-3}^{\frac{1}{3}}$}\\
  \hline
 $3.\nu_{\rm c}<\nu_{\rm m}<\nu_{\rm opt}$& -- &\tabincell{c}{$\Gamma_{0,2}< 3.0 \nu_{\rm opt,R}^{\frac{1}{18}}n_0^{-\frac{5}{36}}\bar{\epsilon}_{\rm e,-1}^{-\frac{1}{9}}$\\ $\epsilon_{\rm B,-2}^{-\frac{1}{36}}(1+z)^{\frac{5}{18}}f_{\gamma,-7}^{*\frac{1}{9}}D_{\rm L,28}^{\frac{2}{9}}t_{\oplus}^{-\frac{1}{3}}$} &\tabincell{c}{$\Gamma_{0,2}>4.2 n_0^{-\frac{2}{11}}\bar{\epsilon}_{\rm e,-1}^{-\frac{1}{11}}\epsilon_{\rm B,-2}^{-\frac{1}{11}}(1+Y)^{-\frac{1}{11}}$\\ $(1+z)^{\frac{3}{11}}f_{\gamma,-7}^{*\frac{1}{11}}D_{\rm L,28}^{\frac{2}{11}}t_{\oplus}^{-\frac{4}{11}}$} & \tabincell{c}{$\Gamma_{0,2}<2.2\nu_{\rm opt,R}^{\frac{5}{66}}n_0^{-\frac{17}{132}}\bar{\epsilon}_{\rm e,-1}^{-\frac{1}{11}}$\\ $\epsilon_{\rm B,-2}^{-\frac{1}{132}}(1+Y)^{\frac{2}{33}}(1+z)^{\frac{17}{66}} $\\ $f_{\gamma,-7}^{*\frac{2}{33}}D_{\rm L,28}^{\frac{8}{33}}t_{\oplus}^{-\frac{10}{33}}F_{\rm \nu,limit,-3}^{\frac{2}{33}}$}\\
      \hline
 $4.\nu_{\rm opt}<\nu_{\rm m}<\nu_{\rm c}$ & --& \tabincell{c}{$\Gamma_{0,2}> 3.0 \nu_{\rm opt,R}^{\frac{1}{18}}n_0^{-\frac{5}{36}}\bar{\epsilon}_{\rm e,-1}^{-\frac{1}{9}}$\\ $\epsilon_{\rm B,-2}^{-\frac{1}{36}}(1+z)^{\frac{5}{18}}f_{\gamma,-7}^{*\frac{1}{9}}D_{\rm L,28}^{\frac{2}{9}}t_{\oplus}^{-\frac{1}{3}}$}& \tabincell{c}{$\Gamma_{0,2}<4.2 n_0^{-\frac{2}{11}}\bar{\epsilon}_{\rm e,-1}^{-\frac{1}{11}}\epsilon_{\rm B,-2}^{-\frac{1}{11}}(1+Y)^{-\frac{1}{11}}$\\ $(1+z)^{\frac{3}{11}}f_{\gamma,-7}^{*\frac{1}{11}}D_{\rm L,28}^{\frac{2}{11}}t_{\oplus}^{-\frac{4}{11}}$} &\tabincell{c}{$\Gamma_{0,2}>1.5 \times 10^2\nu_{\rm opt,R}^{\frac{1}{3}}n_0^{\frac{1}{6}}$\\ $\bar{\epsilon}_{\rm e,-1}^{-\frac{2}{3}}\epsilon_{\rm B,-2}^{\frac{1}{3}}(1+z)^{\frac{2}{3}}f_{\gamma,-7}^{*\frac{7}{6}}$\\ $ D_{\rm L,28}^{\frac{1}{3}}t_{\oplus}^{-\frac{1}{2}}F_{\rm \nu,limit,-3}^{-1}$}\\
 \hline
 $5.\nu_{\rm m}<\nu_{\rm opt}<\nu_{\rm c}$ & \tabincell{c}{$\Gamma_{0,2} <20.2 \nu_{\rm opt,R}^{-\frac{1}{4}}n_0^{-\frac{3}{8}}\epsilon_{\rm B,-2}^{-\frac{3}{8}}$\\ $(1+Y)^{-\frac{1}{2}}(1+z)^{\frac{1}{4}}t_{\oplus}^{-\frac{1}{2}}$} &  \tabincell{c}{$\Gamma_{0,2}< 3.0 \nu_{\rm opt,R}^{\frac{1}{18}}n_0^{-\frac{5}{36}}\bar{\epsilon}_{\rm e,-1}^{-\frac{1}{9}}$\\ $\epsilon_{\rm B,-2}^{-\frac{1}{36}}(1+z)^{\frac{5}{18}}f_{\gamma,-7}^{*\frac{1}{9}}D_{\rm L,28}^{\frac{2}{9}}t_{\oplus}^{-\frac{1}{3}}$}& --& \tabincell{c}{$\Gamma_{0,2}<2.8 \nu_{\rm opt,R}^{\frac{3}{74}}n_0^{-\frac{23}{148}}\bar{\epsilon}_{\rm e,-1}^{-\frac{3}{37}}$\\ $\epsilon_{\rm B,-2}^{-\frac{7}{148}}(1+z)^{\frac{19}{74}}f_{\gamma,-7}^{*\frac{2}{37}}$\\ $D_{\rm L,28}^{\frac{8}{37}}t_{\oplus}^{-\frac{12}{37}}F_{\rm \nu,limit,-3}^{*\frac{2}{37}}$}\\
 \hline
 $6.\nu_{\rm m}<\nu_{\rm c} <\nu_{\rm opt}$	&	\tabincell{c}{$\Gamma_{0,2}>20.2 \nu_{\rm opt,R}^{-\frac{1}{4}}n_0^{-\frac{3}{8}}\epsilon_{B\rm ,-2}^{-\frac{3}{8}}$\\ $(1+Y)^{-\frac{1}{2}}(1+z)^{\frac{1}{4}}t_{\oplus}^{-\frac{1}{2}}$} &--& \tabincell{c}{$\Gamma_{0,2}<4.2 n_0^{-\frac{2}{11}}\bar{\epsilon}_{\rm e,-1}^{-\frac{1}{11}}\epsilon_{\rm B,-2}^{-\frac{1}{11}}(1+Y)^{-\frac{1}{11}}$\\ $(1+z)^{\frac{3}{11}}f_{\gamma,-7}^{*\frac{1}{11}}D_{\rm L,28}^{\frac{2}{11}}t_{\oplus}^{-\frac{4}{11}}$}&\tabincell{c}{$\Gamma_{0,2}<2.2\nu_{\rm opt,R}^{\frac{5}{66}}n_0^{-\frac{17}{132}}\bar{\epsilon}_{\rm e,-1}^{-\frac{1}{11}}$\\ $\epsilon_{\rm B,-2}^{-\frac{1}{132}}(1+Y)^{\frac{2}{33}}(1+z)^{\frac{17}{66}} $\\ $f_{\gamma,-7}^{*\frac{2}{33}}D_{\rm L,28}^{\frac{8}{33}}t_{\oplus}^{-\frac{10}{33}}F_{\rm \nu,limit,-3}^{\frac{2}{33}}$}\\
 \hline
 \end{tabular}}
\end{table*}
%
%
\begin{table*}
\centering
\caption{The constraints on the initial Lorentz factor $\Gamma_0$ in the NRS case for a wind environment.}
\label{ThinW}
\resizebox{\textwidth}{!}{ %
 \begin{tabular}{|c|c|c|c|c|}
   \hline &
 ($\nu_{\rm opt}$, $\nu_{\rm c}$) & ($\nu_{\rm opt}$, $\nu_{\rm m}$) & ($\nu_{\rm m}$, $\nu_{\rm c}$) &$F_{\nu} < F_{\nu,limit}$ \\
   \hline
 $1.\nu_{\rm opt}<\nu_{\rm c}<\nu_{\rm m}$	&	\tabincell{c}{$\Gamma_{0,2} >12.1 \nu_{\rm opt,R}^{\frac{1}{2}}A_{*,-1}^{\frac{3}{4}}$\\ $\epsilon_{\rm B,-2}^{\frac{3}{4}}(1+Y)(1+z)t_{\oplus}^{-\frac{1}{2}}$}	&	 -- &\tabincell{c}{$\Gamma_{0,2}>0.1 A_{*,-1}^{-\frac{2}{3}}\bar{\epsilon}_{\rm e,-1}^{-\frac{1}{3}}\epsilon_{\rm B,-2}^{-\frac{1}{3}}$\\ $(1+Y)^{-\frac{1}{3}}(1+z)^{-\frac{1}{3}}f_{\gamma,-7}^{*\frac{1}{3}}D_{\rm L,28}^{\frac{2}{3}}$}		& \tabincell{c}{$\Gamma_{0,2}<2.2 \times 10^{-15}\nu_{\rm opt,R}^{-1}A_{*,-1}^{-\frac{9}{2}}\epsilon_{\rm B,-2}^{-3}$\\ $(1+Y)^{-2}(1+z)^{-5}f_{\gamma,-7}^{*-\frac{3}{2}}D_{\rm L,28}^{3}t_{\oplus}^{\frac{5}{2}}F_{\rm \nu,limit,-3}^3 $}\\
   \hline
 $2. \nu_{\rm c}<\nu_{\rm opt}<\nu_{\rm m}$ & \tabincell{c}{$\Gamma_{0,2} <12.1 \nu_{\rm opt,R}^{\frac{1}{2}}A_{*,-1}^{\frac{3}{4}}$\\ $\epsilon_{\rm B,-2}^{\frac{3}{4}}(1+Y)(1+z)t_{\oplus}^{-\frac{1}{2}}$}&  \tabincell{c}{$\Gamma_{0,2} > 0.3\nu_{\rm opt,R}^{\frac{1}{8}} A_{*,-1}^{-\frac{5}{16}}\bar{\epsilon}_{\rm e,-1}^{-\frac{1}{4}}$\\ $\epsilon_{\rm B,-2}^{-\frac{1}{16}}f_{\gamma,-7}^{*\frac{1}{4}}D_{\rm L,28}^{\frac{1}{2}}t_{\oplus}^{-\frac{1}{8}}$}& --& \tabincell{c}{$\Gamma_{0,2}<2.8 \times 10^{-2}\nu_{\rm opt,R}^{\frac{1}{4}}A_{*,-1}^{-\frac{1}{8}}\epsilon_{\rm B,-2}^{\frac{1}{8}}$\\ $(1+Y)^{\frac{1}{2}} f_{\gamma,-7}^{*-\frac{1}{4}}D_{\rm L,28}^{\frac{1}{2}}F_{\rm \nu,limit,-3}^{\frac{1}{2}}$}\\
 \hline
 $3.\nu_{\rm c}<\nu_{\rm m}<\nu_{\rm opt}$ & --& \tabincell{c}{$\Gamma_{0,2} < 0.3\nu_{\rm opt,R}^{\frac{1}{8}} A_{*,-1}^{-\frac{5}{16}}\bar{\epsilon}_{\rm e,-1}^{-\frac{1}{4}}$\\ $\epsilon_{\rm B,-2}^{-\frac{1}{16}}f_{\gamma,-7}^{*\frac{1}{4}}D_{\rm L,28}^{\frac{1}{2}}t_{\oplus}^{-\frac{1}{8}}$}& \tabincell{c}{$\Gamma_{0,2}>0.1 A_{*,-1}^{-\frac{2}{3}}\bar{\epsilon}_{\rm e,-1}^{-\frac{1}{3}}\epsilon_{\rm B,-2}^{-\frac{1}{3}}$\\ $(1+Y)^{-\frac{1}{3}}(1+z)^{-\frac{1}{3}}f_{\gamma,-7}^{*\frac{1}{3}}D_{\rm L,28}^{\frac{2}{3}}$}		 &\tabincell{c}{$\Gamma_{0,2}<0.2 \nu_{\rm opt,R}^{\frac{5}{32}}A_{*,-1}^{-\frac{17}{64}} \bar{\epsilon}_{\rm e,-1}^{-\frac{3}{16}}\epsilon_{\rm B,-2}^{-\frac{1}{64}}$\\ $(1+Y)^{\frac{1}{8}}f_{\gamma,-7}^{*\frac{1}{8}}D_{\rm L,28}^{\frac{1}{2}} t_{\oplus}^{-\frac{3}{32}}F_{\rm \nu,limit,-3}^{\frac{1}{8}}$}\\
 \hline
 $4. \nu_{\rm opt}<\nu_{\rm m}<\nu_{\rm c}$& --& \tabincell{c}{$\Gamma_{0,2} > 0.3\nu_{\rm opt,R}^{\frac{1}{8}} A_{*,-1}^{-\frac{5}{16}}\bar{\epsilon}_{\rm e,-1}^{-\frac{1}{4}}$\\ $\epsilon_{\rm B,-2}^{-\frac{1}{16}}f_{\gamma,-7}^{*\frac{1}{4}}D_{\rm L,28}^{\frac{1}{2}}t_{\oplus}^{-\frac{1}{8}}$} &\tabincell{c}{$\Gamma_{0,2}<0.1 A_{*,-1}^{-\frac{2}{3}}\bar{\epsilon}_{\rm e,-1}^{-\frac{1}{3}}\epsilon_{\rm B,-2}^{-\frac{1}{3}}$\\ $(1+Y)^{-\frac{1}{3}}(1+z)^{-\frac{1}{3}}f_{\gamma,-7}^{*\frac{1}{3}}D_{\rm L,28}^{\frac{2}{3}}$} &\tabincell{c}{$\Gamma_{0,2} >55.7\nu_{\rm opt,R}^{\frac{1}{5}}A_{*,-1}^{\frac{1}{10}}\bar{\epsilon}_{\rm e,-1}^{-\frac{2}{5}}\epsilon_{\rm B,-2}^{\frac{1}{5}}$\\ $(1+z)^{\frac{3}{5}}f_{\gamma,-7}^{*\frac{7}{10}}D_{\rm L,28}^{\frac{1}{5}}t_{\oplus}^{-\frac{1}{2}}F_{\rm \nu,limit,-3}^{-\frac{3}{5}}$}\\
   \hline
 $5.\nu_{\rm m}<\nu_{\rm opt}<\nu_{\rm c}$ & \tabincell{c}{$\Gamma_{0,2} >12.1 \nu_{\rm opt,R}^{\frac{1}{2}}A_{*,-1}^{\frac{3}{4}}$\\ $\epsilon_{\rm B,-2}^{\frac{3}{4}}(1+Y)(1+z)t_{\oplus}^{-\frac{1}{2}}$} &\tabincell{c}{$\Gamma_{0,2} < 0.3\nu_{\rm opt,R}^{\frac{1}{8}} A_{*,-1}^{-\frac{5}{16}}\bar{\epsilon}_{\rm e,-1}^{-\frac{1}{4}}$\\ $\epsilon_{\rm B,-2}^{-\frac{1}{16}}f_{\gamma,-7}^{*\frac{1}{4}}D_{\rm L,28}^{\frac{1}{2}}t_{\oplus}^{-\frac{1}{8}}$}& --		& \tabincell{c}{$\Gamma_{0,2}< 0.1 \nu_{\rm opt,R}^{\frac{3}{28}}A_{*,-1}^{-\frac{23}{56}}\bar{\epsilon}_{\rm e,-1}^{-\frac{3}{14}}\epsilon_{\rm B,-2}^
 {-\frac{1}{8}}$\\$(1+z)^{-\frac{1}{7}}f_{\gamma,-7}^{*\frac{1}{7}}D_{\rm L,28}^{\frac{4}{7}}t_{\oplus}^{-\frac{1}{28}}F_{\rm \nu,limit,-3}^{\frac{1}{7}}$}\\
 \hline
 $6.\nu_{\rm m}<\nu_{\rm c}<\nu_{\rm opt}$ &\tabincell{c}{$\Gamma_{0,2} <12.1 \nu_{\rm opt,R}^{\frac{1}{2}}A_{*,-1}^{\frac{3}{4}}$\\ $\epsilon_{\rm B,-2}^{\frac{3}{4}}(1+Y)(1+z)t_{\oplus}^{-\frac{1}{2}}$}	&	 -- &\tabincell{c}{$\Gamma_{0,2}<0.1 A_{*,-1}^{-\frac{2}{3}}\bar{\epsilon}_{\rm e,-1}^{-\frac{1}{3}}\epsilon_{\rm B,-2}^{-\frac{1}{3}}$\\ $(1+Y)^{-\frac{1}{3}}(1+z)^{-\frac{1}{3}}f_{\gamma,-7}^{*\frac{1}{3}}D_{\rm L,28}^{\frac{2}{3}}$}&\tabincell{c}{$\Gamma_{0,2}<0.2 \nu_{\rm opt,R}^{\frac{5}{32}}A_{*,-1}^{-\frac{17}{64}} \bar{\epsilon}_{\rm e,-1}^{-\frac{3}{16}}\epsilon_{\rm B,-2}^{-\frac{1}{64}}$\\ $(1+Y)^{\frac{1}{8}}f_{\gamma,-7}^{*\frac{1}{8}}D_{\rm L,28}^{\frac{1}{2}} t_{\oplus}^{-\frac{3}{32}}F_{\rm \nu,limit,-3}^{\frac{1}{8}}$}\\
 \hline
 \end{tabular}}%
\end{table*}
\subsection{Data Analysis}
A large sample of ROTSE-III optical data reported from February 2005 to July 2011 is collected. There is a total of 62 GRBs including 18 detections and 44 lower-limit measurements of magnitude. The shape of the most of the detected lightcurves are complex that it is difficult to identify the reverse-shock component. Thus we only consider the lower-limit sample in this work. From Table \ref{Thick} and Table \ref{ThickW}, we find that the constraints on the RS model with the Lorentz factor $\Gamma_0$ for the RRS case can be obtained for the bursts with four observables, i.e. the redshift of GRB $z$, the duration of prompt gamma-ray emission $T_{90}$, the fluence in gamma-ray band $f_{\gamma}$ and the optical limit $F_{\rm \nu,limit}$ around the time of $T_{90}$. A fraction of the GRBs in this sample have higher attenuations from their host galaxies, for example, GRB 071025 (Perley et al. 2010), GRB 100621A (Greiner et al. 2013), GRB 061222A and GRB 090709A (Perley et al. 2013). We exclude those bursts with extinction $A_{\rm v}>2$ in their host galaxies from previous studies (e.g., Cenko et al. 2009; Greiner et al. 2011; Perley et al. 2013) and adopt 36 lower-limit measurements with the needed observables. The properties of prompt emissions including the duration and the fluence of these bursts are taken from {\it Swift} GRB reports\footnote{http://swift.gsfc.nasa.gov/archive/}. We further collect the redshifts of GRBs in our sample from Jochen Greiner's Table\footnote{http://www.mpe.mpg.de/$\sim$jcg/grbgen.html}. For the bursts in the sample without redshift measurements, we assume redshifts of $z=2$ for the calculation since the mean redshift of {\it Swift} GRBs has been shown to be close to 2 (e.g., Fynbo et al. 2009). The lower-limit magnitude measured near the time of $T_{90}$ is selected and transformed into the observed upper-limit of flux $F_{\rm \nu,limit}$. In the thin shell case, the constraints are achieved by the inequalities presented in Table \ref{Thin} and Table \ref{ThinW}. The observed data after the end of the duration $T_{90}$ are considered and the lower-limit magnitude is determined by the most conservative estimation described above. We collect 26 bursts with the limit number of points no less than five after the end of the duration $T_{90}$ from our lower-limits GRB sample. Most of the GRBs in this sample are the same as those selected for the RRS case except one burst GRB 090621A, which is without the information about $T_{90}$ and $f_{\gamma}$ in the database of {\it Swift}/BAT. Two observables including the redshift $z$ and fluence $f_{\gamma}$ in prompt gamma-ray emission are needed for the GRBs in this sample to constrain the initial Lorentz factor once the lower-limit magnitude is obtained. A fluence of $f_{\gamma}=10^{-7}$ erg cm$^{-2}$ is assumed for the burst GRB 090621A in our calculations.

The data processing for ROTSE-III are similar to the procedures presented in the work of Cui et al. (2014) with some minor changes and are summarized here. In the work of Cui et al. (2014), for uniformity purpose when studying the GRB luminosity function, all observations were interpolated or extrapolated to a common time at 100s after the burst, while here we do not apply this procedure. But similar to Cui et al. (2014), since most of the lower limit measurements were obtained with an exposure of 5s, we normalized all the longer exposure times (either 20s or 60s) to 5s exposures. A transformation offset magnitude was applied to the lower-limit measurements of longer exposure time when transforming into the 5s-exposure lower limit. Since the signal-to-noise ratio is proportional to the exposure time with power index of 0.5, the offset magnitude was calculated as $\Delta m = -2.5 \log [(\frac{t_{exp}}{5})^{0.5}]$, where $t_{exp}$ is the longer exposure time, either 20s or 60s. For example, if the limit magnitude is 17.7 mag for 20s exposure, the corresponding 5s limit magnitude is $\sim$17.0 mag after applying the offset mag of -0.7 mag. Since all the ROTSE-III observations were taken in an unfiltered band, which is roughly equivalent to the $R_{\rm C}$ band system (Rykoff et al. 2009), the magnitudes are converted to flux by assuming they are equal to $R_{\rm C}$ magnitude. The estimation of extinction in GRB host galaxy for the individual burst involves in the effects of redshift and the change of attenuation law. Since it is difficult to quantify the exact values of these effects, we apply the same extinction correction procedure in Milky Way galaxy (Schlafly \& Finkbeiner 2011) and in GRB host galaxy (the mean value of $A_{\rm v}$=0.2, Kann et al. 2010) as those given in Cui et al. (2014).

In the homogeneous ISM, a critical initial Lorentz factor $\Gamma_{\rm c}$ is defined as (Zhang, Kobayashi, \& M{\'e}sz{\'a}ros 2003)
\begin{equation}
\label{Gacri}
\Gamma_{\rm c,ISM} \simeq 96 E_{52}^{1/8} n_0^{-1/8} T_{90,2}^{-3/8}(1+z)^{3/8}
\end{equation}
if the burst duration is equal to the deceleration time of the NRS when the mass of ISM accumulated by the FS is $m_{\rm ISM}=m_{\rm ej}/\Gamma_0$, where $m_{\rm ej}$ is the ejecta mass. In wind environment, the critical Lorentz factor is (Zou, Wu, \& Dai 2005)
\begin{equation}
\label{Gacriw}
\Gamma_{\rm c,wind} \simeq 41 E_{52}^{1/4} A_{*,-1}^{-1/4} T_{90,2}^{-1/4}(1+z)^{1/4}.
\end{equation}
For the thick-shell case, the initial Lorentz factor should satisfy $\Gamma_0>\Gamma_{\rm c}$; and for thin shells, $\Gamma_0<\Gamma_{\rm c}$. Here, $\Gamma_{\rm c}=\Gamma_{\rm c,ISM}$ or $\Gamma_{\rm c,wind}$ for ISM or wind environment, respectively. This can give additional constraints on the initial Lorentz factor achieved from the upper limits of the ROTSE-III sample. And we only consider the limits of $\Gamma_0$ in the regimes of 10 to 10$^4$ in this work.

With what follows, we present results for energy fraction $\epsilon_{\rm e}=0.1$ and $\epsilon_{\rm B}=0.01$, as well as the electron power law index $p$=2.5 such that the time-integrated synchrotron emission is in agreement with the observations (Panaitescu \& Kumar 2001, 2002). The investigations about the electron index suggested that the distribution of $p$ is likely a Gaussian distribution from 2 to 3.5 with a typical value 2.5 (e.g., Liang et al. 2013; Wang et al. 2015). The parameters for density are adopted as $n_0$=1 for standard homogenous ISM and $A_*=0.1$ for the wind. These values are the typical values obtained from multi-wavelengths to the afterglow spectra (Panaitescu \& Kumar 2001, 2002). In the early FS-RS processes, the inverse Compton losses are small and the synchrotron cooling is dominated thus the Compton parameter $Y < 1$ in the RS region. This assumption was also adopted and the factor $1+Y$ was neglected in the previous studies (e.g., Kumar 2000; Lloyd-Ronning \& Zhang 2004; D'Avanzo et al. 2012). The radiative efficiency $\eta_{\rm \gamma}$ is to probe how efficient a GRB converts its total energy of the central engine into prompt $\gamma$-ray emission. But its value was found to be highly model-dependent and varies in the wide range of 0.01 to 0.9 (e.g., Kobayashi \& Sari 2001; Guetta, Spada \& Waxman 2001). Here, we assume the efficiency fraction $\eta_\gamma=0.2$ in our calculations. In reality, the ranges of the microphysical parameters are large among different bursts and they may even vary with time for individual burst. We discuss the effect of the density parameter $n_0$ and $A_*$ on $\Gamma_0$ in the third paragraph of Section 4 after considering $n_0$ in the regime of [$10^{-3}$, $10^3$] and $A_*$ in [$10^{-3}$, 10].

\section{Results}
\begin{figure*}
	\includegraphics[width=8.0cm]{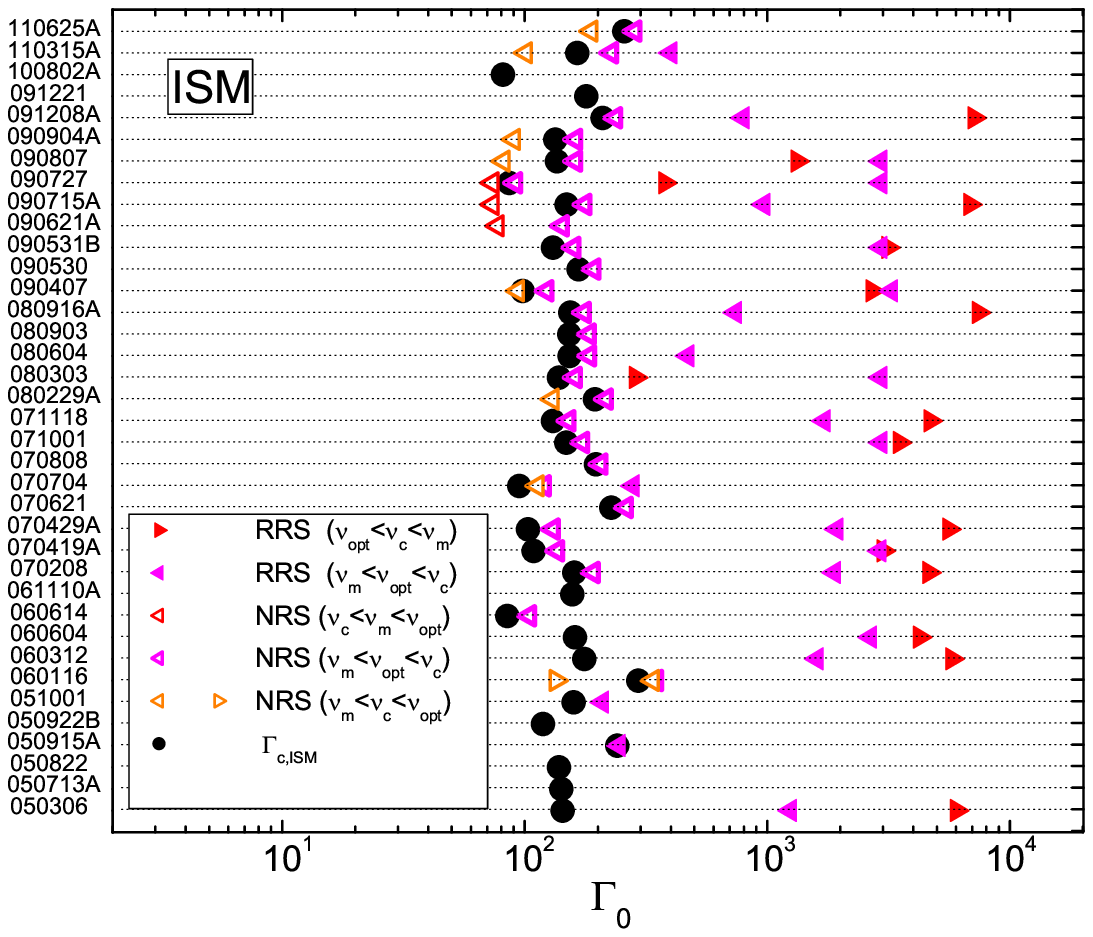}
    \includegraphics[width=8.0cm]{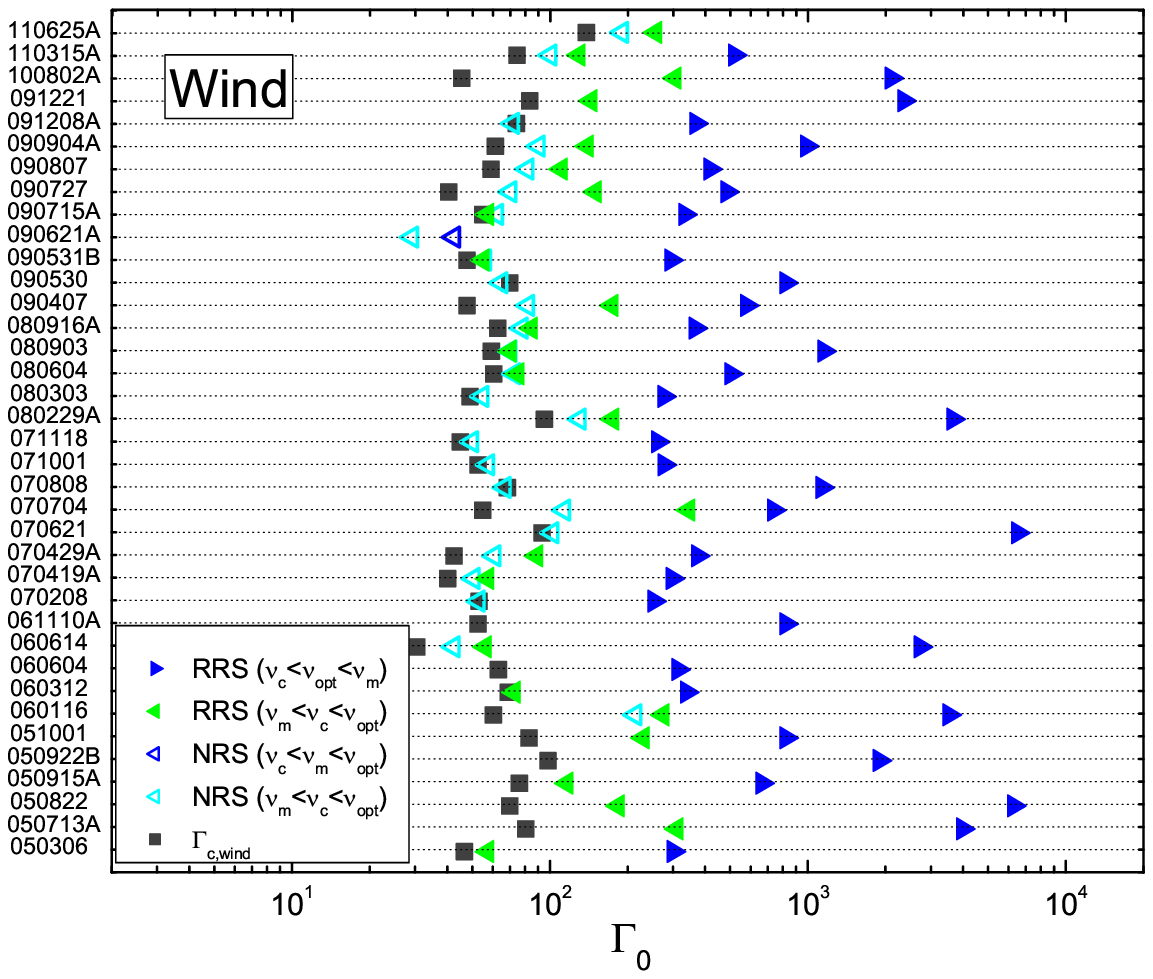}
    \caption{The RS model constraints and the estimates of the initial Lorentz factor $\Gamma_0$ using the models described in Section 2.1. Left and right panels are for the homogeneous ISM and wind environment, respectively. The types of triangles correspond to limits of $\Gamma_0$ in the different spectral regimes for the NRS or the RRS cases. The right-arrowed triangles denote the lower limits of $\Gamma_0$, while the left-arrowed ones denote the upper limits. Dark solid circles and squares correspond to the critical Lorentz factors in ISM and wind, respectively.}
    \label{Gammadi}
\end{figure*}

The properties of the GRBs except GRB 090621A in our sample including the duration $T_{90}$, the prompt gamma-ray fluence $f_{\gamma}$, the optical limit of flux at about $T_{90}$, $F_{\rm \nu,limit}$, the critical Lorentz factor $\Gamma_{\rm c,ISM}$ and $\Gamma_{\rm c,wind}$ are presented in Table \ref{result}. The duration $T_{90}$ and the fluence $f_{\gamma}$ are from the catalogue of {\it Swift} GRB table. Based on the above analysis in \S~2, we achieve the constraints on RS models with the initial Lorentz factor $\Gamma_0$ for the bursts in our samples. We show the results in the left and right panels of Figure \ref{Gammadi} for the homogeneous ISM and wind environment, respectively.
\begin{table}[H]
\centering
\caption{The prompt emission properties and the critical Lorentz factors of 40 GRBs in our sample.}
\label{result}
 \begin{tabular}{|c|c|c|c|c|c|}
    \hline
 GRB & $T_{90}$ & $f_{\gamma}$            & $F_{\rm \nu,limit}$ & $\Gamma_{\rm c,ISM}$ & $\Gamma_{\rm c,wind}$ \\
     & (s)      &($10^{-7}$erg cm$^{-2}$) &($10^{-3}$Jy)& &   \\
\hline 
050306	&	158.3	&	115.0		&	0.88	&	126	&	46	\\
050713A	&	124.7	&	51.1	&	1.36	&	143	&	80	\\
050822$\dag$	&	103.4	&	24.6	&	0.38	&	141	&	70	\\
050915A$\dag$	&	52	&	8.5	&	1.64	&	163	&	76	\\
050922B	&	150.9	&	22.3	&	0.45	&	241	&	98	\\
051001$\dag$	&	189	&	17.4	&	0.94	&	159	&	83	\\
060116	&	105.9	&	24.1	&	0.78	&	137	&	60	\\
060312	&	50.6	&	19.7	&	5.42	&	176	&	69	\\
060604$\dag$	&	95.0	&	4.0	&	4.5	&	167	&	63	\\
060614$\dag$	&	108.7	&	204	&	1.64	&	85	&	30	\\
061110A$\dag$	&	40.7	&	10.6	&	0.78	&	157	&	52	\\
070208$\dag$	&	47.7	&	4.45	&	1.03	&	160	&	53	\\
070419A$\dag$	&	115.6	&	5.58	&	1.03	&	109	&	40	\\
070429A	&	163.3	&	9.1	&	0.94	&	103	&	42	\\
070621	&	33.3	&	43.0	&	0.78	&	227	&	93	\\
070704	&	380	&	59	&	1.97	&	95	&	55	\\
070808	&	32	&	12	&	0.94	&	197	&	68	\\
071001	&	58.5	&	7.7	&	2.16	&	148	&	52	\\
071118	&	71	&	5.0	&	0.78	&	131	&	45	\\
080229A	&	64	&	90	&	2.16	&	195	&	95	\\
080303	&	67	&	6.6	&	19.66	&	138	&	49	\\
080604$\dag$	&	82	&	8.0	&	0.78	&	153	&	60	\\
080903	&	66	&	14	&	0.65	&	153	&	59	\\
080916A$\dag$	&	60	&	40	&	6.51	&	154	&	63	\\
090407$\dag$	&	310	&	11	&	2.84	&	98	&	48	\\
090530$\dag$	&	48	&	11	&	0.86	&	185	&	70	\\
090531B	&	80	&	7.1	&	1.79	&	131	&	48	\\
090715A	&	63	&	9.8	&	1.49	&	149	&	55	\\
090727	&	302	&	14	&	17.93	&	86	&	40	\\
090807	&	140.8	&	22	&	17.93	&	136	&	59	\\
090904A	&	122	&	30	&	1.24	&	133	&	61	\\
091208A	&	29.1	&	15	&	4.11	&	209	&	74	\\
091221	&	68.5	&	57	&	1.79	&	180	&	83	\\
100802A	&	487	&	36	&	0.26	&	81	&	45	\\
110315A	&	77	&	41	&	4.94	&	165	&	74	\\
110625A	&	44.5	&	280	&	1.49	&	258	&	138	\\
\hline
\end{tabular}
\begin{tablenotes}
  \item[$\dag$]$\dag$ GRBs with measured redshifts
   \end{tablenotes}
\end{table}

We summarize the constraints on external RS models with the initial Lorentz factor $\Gamma_0$ we achieved in Table \ref{constr}.
\begin{table*}
\centering
\caption{The results of constraints on the RS models with the initial Lorentz factor $\Gamma_0$ in the RRS and the NRS cases.}
\label{constr}
\begin{tabular}{|c|c|c|c|c|c|c|}
 \hline
 RS type&	Environment	&	Spectral regime	&	Limit type	&Bursts&	Constraint  range	\\
 \hline
 \multirow{4}{*}{RRS} & \multirow{2}{*}{ISM} & $\nu_{\rm opt}<\nu_{\rm c}<\nu_{\rm m}$ & lower & 16 & [286, 7421] \\
 \cline{3-6}
 &&$\nu_{\rm m}<\nu_{\rm opt}<\nu_{\rm c}$& upper & 21 & [208, 3239] \\
 \cline{2-6}
   & \multirow{2}{*}{Wind} & $\nu_{\rm c}<\nu_{\rm opt}<\nu_{\rm m}$ &lower& 35 & [252, 6514]  \\
   \cline{3-6}
 &&$\nu_{\rm m}<\nu_{\rm c}<\nu_{\rm opt}$&upper& 25 & [55, 343] \\
 \hline
 \multirow{6}{*}{NRS} & \multirow{4}{*}{ISM} & $\nu_{\rm c}<\nu_{\rm m}<\nu_{\rm opt}$&upper& 3 & [73, 77] \\
 \cline{3-6}
 &&$\nu_{\rm m}<\nu_{\rm opt}<\nu_{\rm c}$&upper& 26 & [91, 349]\\
 \cline{3-6}
 &&\multirow{2}{*}{$\nu_{\rm m}<\nu_{\rm c}<\nu_{\rm opt}$}&upper& 8 & [82, 336] \\
 \cline{4-6}
 &&&lower& 1 & 135\\
 \cline{2-6}
   & \multirow{2}{*}{Wind} & $\nu_{\rm c}<\nu_{\rm m}<\nu_{\rm opt}$ &upper& 1 & 42 \\
   \cline{3-6}
 &&$\nu_{\rm m}<\nu_{\rm c}<\nu_{\rm opt}$&upper& 26 & [29, 213] \\
 \hline
 \end{tabular}
 \end{table*}
From the table, we find that there at least two cases of the spectral regimes can give constraints on each type of the RS (RRS or NRS) with $\Gamma_0$ and each type of medium (ISM or wind) around GRBs. For the constraints on RRS model, we achieve 16 lower limits of the initial Lorentz factor $\Gamma_0$ in the range of $\sim$[290, 7,400] when observed optical frequency $\nu_{\rm opt}$ in the regime of $\nu_{\rm opt}<\nu_{\rm c}<\nu_{\rm m}$ and 21 upper limits in range of $\sim$ [210, 3,240] when $\nu_{\rm m}<\nu_{\rm opt}<\nu_{\rm c}$ for ISM environment. For wind circum-burst medium, there are 35 lower limits in $\sim$[250, 6,500] when $\nu_{\rm c}<\nu_{\rm opt}<\nu_{\rm m}$ and 25 upper limits in about [55, 340] when $\nu_{\rm m}<\nu_{\rm c}<\nu_{\rm opt}$. For the constraints on NRS model, there are total 37 upper limits (in $\sim$[70, 350]) and only one lower limit of $\Gamma_{\rm 0}$ (the constraint regimes of $\nu_{\rm opt}$ showed in the sixth column of Table \ref{constr}) for the homogenous ISM. For wind, there are 27 upper limits distributing among 30 to 210 achieving for two cases, i.e., $\nu_{\rm c}<\nu_{\rm m}<\nu_{\rm opt}$ and $\nu_{\rm m}<\nu_{\rm c}<\nu_{\rm opt}$. All of the upper limits of $\Gamma_{\rm 0}$ achieved in the NRS case are less than 350.

\section{Conclusions and Discussion\label{sec:Conclu}}

In this paper, the constraints on RS models with the initial Lorentz factor $\Gamma_0$ of GRBs are studied based on the assumption that the early optical flash of a GRB is produced by the RS. With the lower limits of magnitude provided by ROTSE-III, we deduce and constrain the values of $\Gamma_0$ for the relativistic RS (thick shell) or for the non-relativistic RS (thin shell). When the observed optical frequency $\nu_{\rm opt}$ in the different regime of minimum synchrotron frequency $\nu_{\rm m}$ and cooling frequency $\nu_{\rm c}$, the upper and lower limits of $\Gamma_0$ can be achieved by the conditions of the peak flux at the shock crossing time less than the upper limit observed by ROTSE-III and the inequalities among frequencies $\nu_{\rm opt}$, $\nu_{\rm m}$ or $\nu_{\rm c}$. We constrain the external RRS model with the initial Lorentz factor $\Gamma_0$ for 36 GRBs reported by ROTSE-III. We achieve 16 lower limits of the initial Lorentz factor $\Gamma_0$ and 21 upper limits in ISM environment. For wind case, there are 35 upper limits and 25 lower limits. Except the case $\nu_{\rm m}<\nu_{\rm c}<\nu_{\rm opt}$ in wind environment with $\Gamma_0$ less than $\sim 340$, the limit range of $\Gamma_0$ is wide in $\sim 210 - 7400$. Almost all constraints on the NRS model with $\Gamma_0$ of 26 bursts are upper limits except one lower limit for wind environment. These limits of $\Gamma_0$ are in the range of $\sim 30 - 350$. The limits of ROTSE-III give the constraints on RS models with the initial Lorentz factors of GRBs. The limits of $\Gamma_0$ achieved for NRS model are less than those for RRS model.

In the standard FS/RS model, the RS can dominate the early optical/IR emission only for very weakly magnetized ejecta. If the original ejecta is magnetized, the emissions from the RS are greatly suppressed and no optical flash will be detectable (see e.g. Mimica et al. 2010). Even in the non-magnetic case, small-scale turbulence generated by the Rayleigh-Taylor instability amplifies
magnetic fields (Duffel \& MacFadyen 2013). The RS emission peaks at a later time and the FS front is unaffected by this instability. Employing the ``mechanical model'' that incorporates the energy conservation, Uhm et al. (2012) showed that the RS light curves exhibit much richer features while the FS light curves are not sensitive to the ejecta stratifications.

\begin{figure*}
\includegraphics[angle=0,scale=0.75]{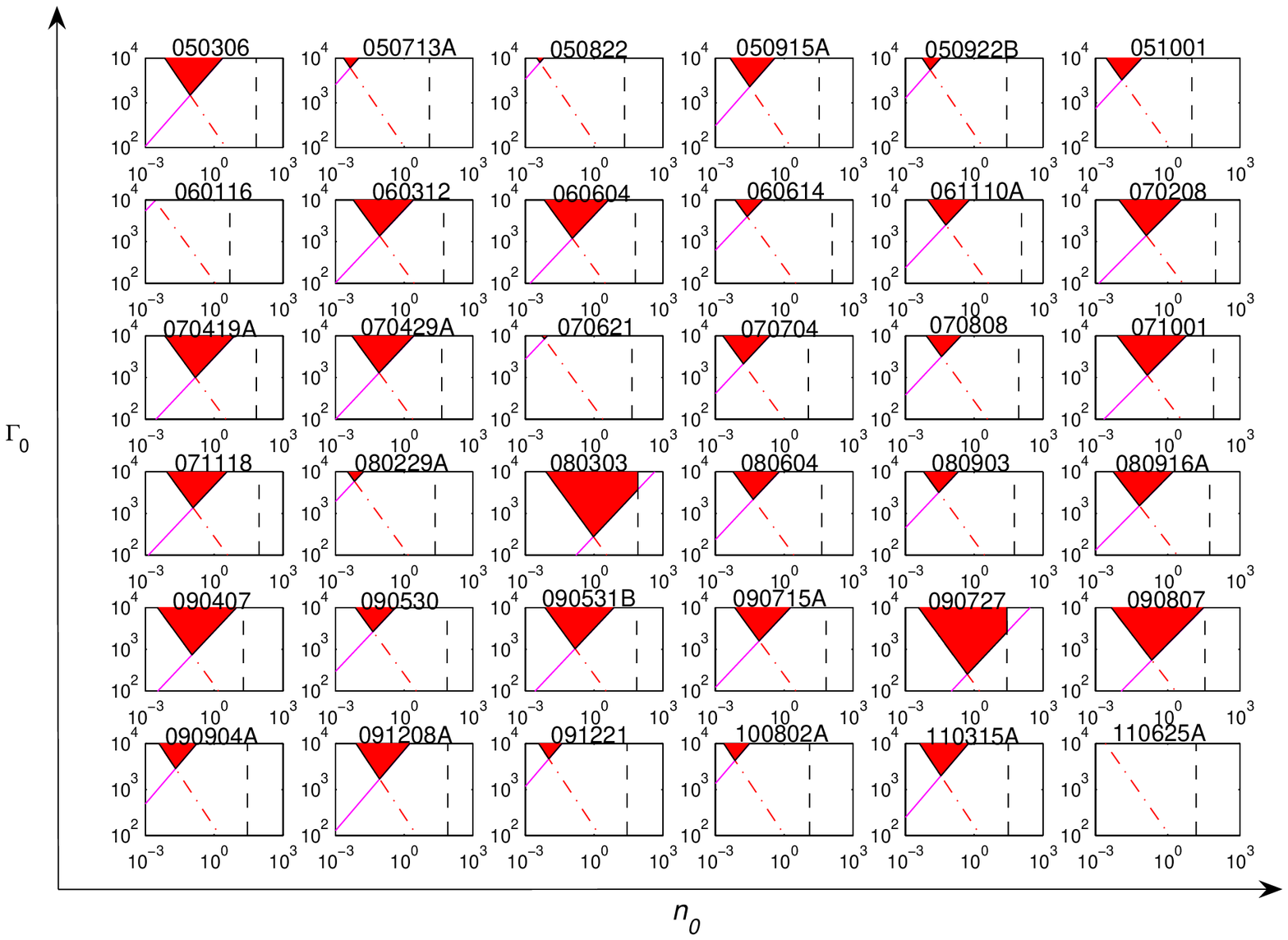}
    \caption{The constraints on the initial Lorentz factor $\Gamma_0$ for the observed frequency $\nu_{\rm obs}$ in the regime of $\nu_{\rm opt}<\nu_{\rm c}<\nu_{\rm m}$ for the RRS case in homogeous ISM environment. The black dash line is the upper limit of density $n_0$ determined by the case $\nu_{\rm opt}<\nu_{\rm c}$. The red dash-dot line corresponds to the lower limit of the Lorentz factor $\Gamma_0$ obtained from the condition $\nu_{\rm c}<\nu_{\rm m}$. The magenta solid line shows the lower limit of $\Gamma_0$ given by $F_{\nu}<F_{\rm \nu, limit}$. The filled red area is the permitted region for $\Gamma_0$ and $n_0$. The title of each panel is the name of GRB analyzed in this case.}
    \label{RRSISM1}
\end{figure*}

\begin{figure*}
    \includegraphics[angle=0,scale=0.75]{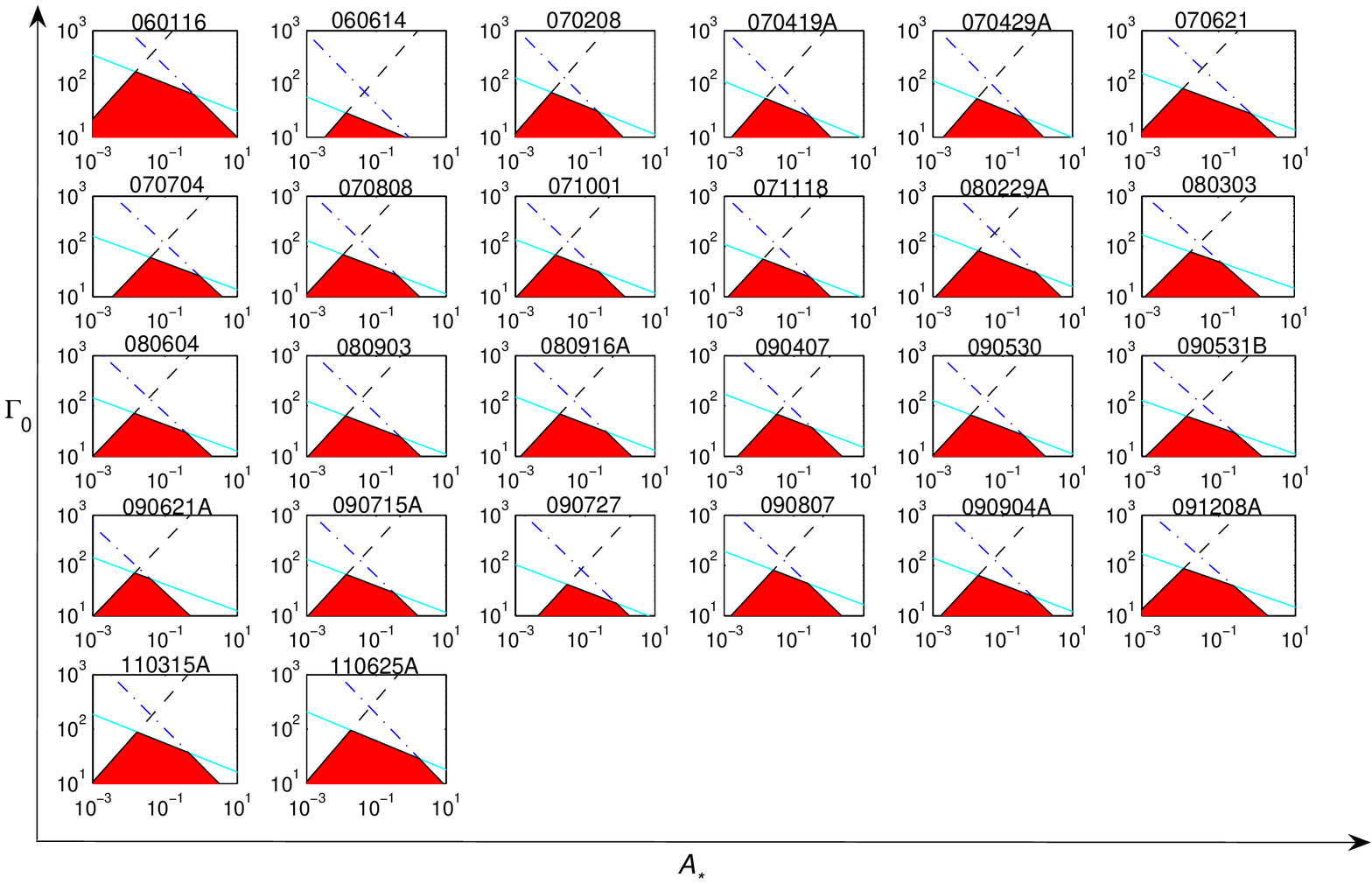}
    \caption{The constraints on the initial Lorentz factor $\Gamma_0$ for the observed frequency $\nu_{\rm obs}$ in the regime of $\nu_{\rm m}<\nu_{\rm c}<\nu_{\rm opt}$ for the NRS case in wind environment. The black dash line is the lower limit of Lorentz factor $\Gamma_0$ obtained from the case $\nu_{\rm c}<\nu_{\rm opt}$. The blue dash-dot line is the lower limit of $\Gamma_0$ determined by the case $\nu_{\rm m}<\nu_{\rm c}$. The cyan solid line shows the upper limits of $\Gamma_0$ given by $F_{\nu}<F_{\rm \nu, limit}$. The filled red area is the permitted region of $\Gamma_0$ and $A_*$. The title of each panel is the name of GRB analyzed in this case.}
    \label{NRSwind6}
\end{figure*}

In order to analyze the effect of density parameters $n_0$ for ISM and $A_*$ for wind environment on the estimation of $\Gamma_0$ limits, we take the values of the parameter $n_0$ in the regime of [$10^{-3}$, $10^3$] and $A_*$ in [$10^{-3}$, 10] to give the constraints on Lorentz factor $\Gamma_0$ based on Table \ref{Thick} to Table \ref{ThinW}. For the NRS case, the principle of the most conservative estimation described above is applied for the limit calculations. We find 9 cases of the spectral regimes for both NRS and RRS models presented in Table \ref{constr} can be constrained with $\Gamma_0$ for most of the bursts in our samples. Figure \ref{RRSISM1} presents one of the examples, corresponding to the spectral regime $\nu_{\rm opt}<\nu_{\rm c}<\nu_{\rm m}$ in ISM environment for the RRS case. Three lines of different color present the constraints on $n_0$ and $\Gamma_0$ given by the three inequalities between ($\nu_{\rm opt}$, $\nu_{\rm c}$), ($\nu_{\rm m}$, $\nu_{\rm c}$) and ($F_{\nu}$, $F_{\rm \nu, limit}$) presented in the second row of Table \ref{Thick}. For example, the black dash line is the upper limits of density parameter $n_0$ obtained from the case $\nu_{\rm opt}< \nu_{\rm c}$. The red dash-dot line corresponds to the lower limit of the Lorentz factor $\Gamma_0$ obtained from case $\nu_{\rm c}<\nu_{\rm m}$. The magenta solid line shows the lower limit of $\Gamma_0$ given by $F_{\nu}<F_{\rm \nu, limit}$. The constraints on Lorentz factor $\Gamma_0$ and density $n_0$ are the common regions among the left part of the black line and upper parts of red and magenta lines (i.e., the filled red area). The title of each panel is the name of GRB analyzed in this case. The GRB without the filled area means that the RRS model in $\nu_{\rm opt}<\nu_{\rm c}<\nu_{\rm m}$ for homogenous ISM environment can't be constrained with the initial Lorentz factor of the burst. From the figure, we can find the limits of $\Gamma_0$ with fixed density parameter $n_0=1$, which is consistent with the results presented in Figure \ref{Gammadi} and Table \ref{constr}. Another example is shown in Figure \ref{NRSwind6} for the case of $\nu_{\rm m}<\nu_{\rm c}<\nu_{\rm opt}$ in wind environment for the NRS case. Three lines show the constraints on parameter $A_*$ and Lorentz factor $\Gamma_0$ given by the three inequalities between ($\nu_{\rm opt}$, $\nu_{\rm c}$), ($\nu_{\rm m}$, $\nu_{\rm c}$) and ($F_{\nu}$, $F_{\rm \nu, limit}$) presented the last row of Table \ref{ThinW}. The filled area is the permitted region of $\Gamma_0$ and $A_*$.
%

\begin{figure*}
    \includegraphics[angle=0,scale=0.85]{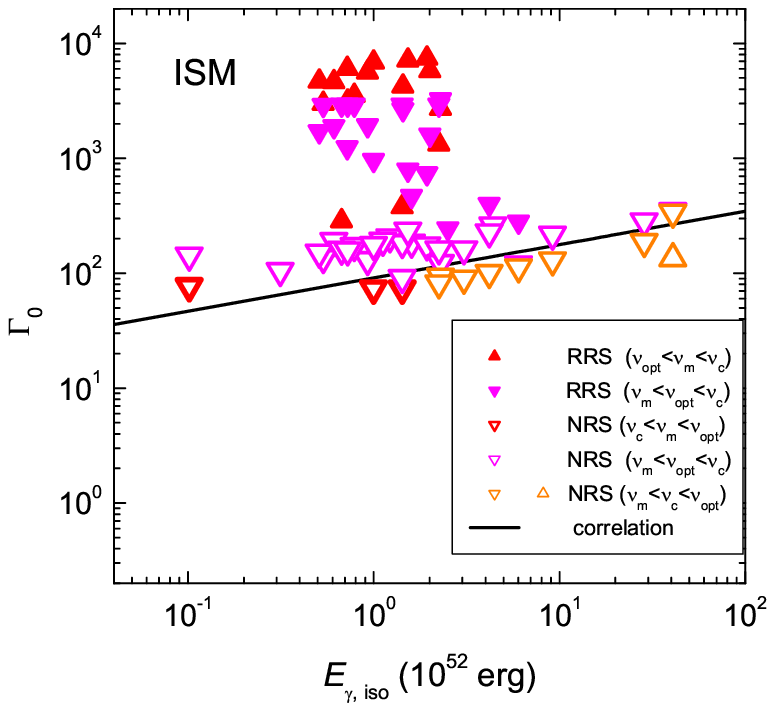}
    \includegraphics[angle=0,scale=0.85]{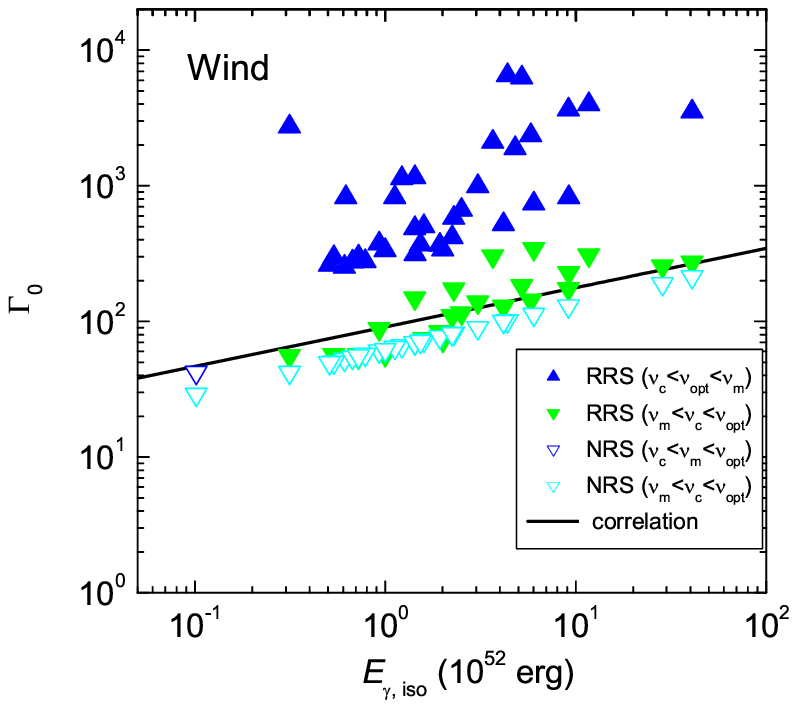}
    \caption{The correlation between the isotropic energy $E_{\rm \gamma, iso}$ in the prompt phase and the initial Lorentz factor $\Gamma_0$ for the GRBs marked with symbol ``$\dag$'' in Table \ref{result} with known redshifts for an ISM (left panel) and a wind (right panel) environment. The solid line is the correlation obtained by L\"u et al. (2012). The upward triangles are lower limits of Lorentz factor $\Gamma_0$ and the inverted ones are upper limits of $\Gamma_0$ deduced from this work. Different colors of triangles denote the same cases as those presented in Figure \ref{Gammadi}.}
    \label{Cor}
\end{figure*}

The correlation between the break time of afterglow plateau and luminosity of GRB has been discussed by van Eerten (2014), who showed that the observed correlations favored thick shell models over thin shell models. In order to account for the observed correlation between the peak energy $E_{\rm p}$ and isotropic energy $E_{\rm \gamma,iso}$ in prompt phase (Amati et al. 2002), the theoretical models demand various correlations between Lorentz factor $\Gamma_0$ and energy $E_{\rm \gamma,iso}$ or luminosity $L_{\rm \gamma,iso}$ (Zhang \& M{\'e}sz{\'a}ros 2002). The correlation between $\Gamma_0$ and $E_{\rm \gamma,iso}$ was discovered by Liang et al. (2010) and was confirmed by Ghirlanda et al. (2012). By updating the GRB sample and more methods to derive $\Gamma_0$, L\"u et al. (2012) confirmed this correlation and found $\Gamma_0 \simeq 91 E_{\rm \gamma, iso, 52}^{0.29}$. There are fifteen GRBs presented with symbol ``$\dag$'' in Table \ref{result} in our optical sample with observed redshifts from Jochen Greiner's Table. We calculate their isotropic energy $E_{\rm \gamma, iso}$ based on observations of {\it Swift}/BAT and plot the correlation between $E_{\rm \gamma, iso}$ and the derived limits of Lorentz factor $\Gamma_0$ in Figure \ref{Cor}. From the figure, we find that though lower limits deduced from thick shell cases are not consistent with the correlation given by L\"u et al. (2012), most of the other limits deduced from ROTSE-III reports are consistent with this correlation. For example, 16 lower limits of $\Gamma_0$ (red solid upward triangles in the left panel) achieved when $\nu_{\rm obs}$ in the regime of $\nu_{\rm opt}<\nu_{\rm m}<\nu_{\rm c}$ for homogenous ISM in RRS case are larger than those predicted from previous relation $\Gamma_0-E_{\rm \gamma, iso, 52}$ (solid line in the panel). There are 35 lower limits of $\Gamma_0$ for RRS case in wind environment (blue solid upward triangles in the right panel), which are larger than predicted ones from previous correlation (solid line). Therefore, the RRS case is not suitable to constrain lower limits of the initial Lorentz factors if we assumed external RS models for the emissions of bursts in our ROTSE III sample.

\section*{Acknowledgements}

We thank the anonymous referee for a constructive report and Z. G. Dai, B. Zhang, L. P. Xin for helpful discussions. This work is partially supported by the National Basic Research Program (``973" Program) of China (Grant Nos 2014CB845800), the National Natural Science Foundation of China (Grants Nos. 11273025, 11433009, 11473038, 11673068, and 11603076), the Youth Innovation Promotion Association (2011231 and 2017366), the Key Research Program of Frontier Sciences (QYZDB-SSW-SYS005) and the Strategic Priority Research Program ``Multi-waveband gravitational wave Universe" (Grant No. XDB23000000) of the Chinese Academy of Sciences, the Natural Science Foundation of Jiangsu Province (Grant No. BK20161096), and the Guangxi Key Laboratory for Relativistic Astrophysics.





\begin{thebibliography}{99}
\bibitem[\protect\citeauthoryear{Abdo}{2009}]{Abdo2009} Abdo, A. A., Ackermann, M., Arimoto, M., et al. 2009, Science, 323, 1688
\bibitem[\protect\citeauthoryear{Ackermann}{2010}]{Ackermann2010} Ackermann, M., Asano, K., Atwood, W. B., et al. 2010, \apj, 716, 1178
\bibitem[\protect\citeauthoryear{Akerlof}{1999}]{Akerlof1999} Akerlof, C. W., Balsano, R., Barthelmy, S., et al. 1999, Nature, 398, 400
\bibitem[\protect\citeauthoryear{Akerlof}{2003}]{Akerlof2003} Akerlof, C. W., Kehoe, R. L., McKay, T. A., et al. 2003, \pasp, 115, 132
\bibitem[\protect\citeauthoryear{Allen}{1973}]{Allen1973} Allen, C. W. Astrophysical Quantities, The Athlone Press, 1973
\bibitem[\protect\citeauthoryear{Amati}{2002}]{Amati2002} Amati, L., Frontera, F., Tavani, M., et al. 2002, A\&A, 390, 81
\bibitem[\protect\citeauthoryear{Aoi}{2010}]{Aoi2010} Aoi, J., Murase, K., Takahashi, K., Ioka, K. \& Nagataki, S. 2010, \apj, 722,440
\bibitem[\protect\citeauthoryear{Baring}{1997}]{Baring1997} Baring, M. G. \& Harding, A. K. 1997, \apj, 491, 663
\bibitem[\protect\citeauthoryear{Bucciantini}{2008}]{Bucciantini2008} Bucciantini, N., Quataert, E., Arons, J., et al. 2008, MNRAS, 383, L25
\bibitem[\protect\citeauthoryear{Bucciantini}{2009}]{Bucciantini2009} Bucciantini, N., Quataert, E., Metzger, B. D., et al. 2009, MNRAS, 396, 2038
\bibitem[\protect\citeauthoryear{Cenko}{2009}]{Cenko2009} Cenko, S. B., Kelemen, J., Harrison, F. A., et al. 2009, \apj, 693, 1484
\bibitem[\protect\citeauthoryear{Chen}{2007}]{Chen2007} Chen, W. X. \& Beloborodov, A. M. 2007, \apj, 657, 383
\bibitem[\protect\citeauthoryear{Chevalier}{1999}]{Chevalier1999} Chevalier, R. A. \& Li, Z. Y. 1999. \apjl, 520, L29
\bibitem[\protect\citeauthoryear{Chevalier}{2000}]{Chevalier2000} Chevalier, R. A. \& Li, Z. Y. 2000, \apj, 536, 195
\bibitem[\protect\citeauthoryear{Cui}{2014}]{Cui2014} Cui, X. H., Wu, X. F., Wei, J. J., et al.\ 2014, \apj, 795, 103
\bibitem[\protect\citeauthoryear{DaiandLu}{1998a}]{DaiandLu1998a} Dai, Z. G. \& Lu, T. 1998a, A\&A 333, L87
\bibitem[\protect\citeauthoryear{Dai}{1998}]{Dai1998} Dai, Z. G. \& Lu, T. 1998b, MNRAS, 298, 87
\bibitem[\protect\citeauthoryear{Dai}{2006}]{Dai2006} Dai, Z. G., Wang, X. Y., Wu, X. F. \& Zhang, B. 2006, Science, 311, 1127
\bibitem[\protect\citeauthoryear{D'Avanzo}{2012}]{D'Avanzo2012} D'Avanzo, P., Salvaterra, R., Sbarufatti, B., et al. 2012, MNRAS, 425, 506
\bibitem[\protect\citeauthoryear{Duffel}{2013}]{Duffel2013} Duffel, P. C. \& MacFadyen, A. I. 2013, \apj, 775, 87
\bibitem[\protect\citeauthoryear{Fenimore}{1993}]{Fenimore1993} Fenimore, E. E., Epstein, R. I. \& Ho, C. 1993, A\&AS, 97, 59
\bibitem[\protect\citeauthoryear{Fynbo}{2009}]{Fynbo2009} Fynbo, J. P., Jakobsson, P., Prochaska, J. X., et al. 2009, \apjs, 185, 526
\bibitem[\protect\citeauthoryear{Gao}{2013}]{Gao2013} Gao, H., Lei, W. H., Zou, Y. C., et al. 2013, NewAR, 57, 141
\bibitem[\protect\citeauthoryear{Ghirlanda}{2012}]{Ghirlanda2012} Ghirlanda, G., Nava, L., Ghisellini, G., et al. 2012, MNRAS, 420, 483
\bibitem[\protect\citeauthoryear{Greiner}{2011}]{Greiner2011} Greiner, J. Kr{\"u}hler, T., Klose, S., et al. 2011, A\&A, 526, 30
\bibitem[\protect\citeauthoryear{Greiner}{2013}]{Greiner2013} Greiner, J., Kr{\"u}hler, T., Nardini, M., et al. 2013, A\&A, 560, 70
\bibitem[\protect\citeauthoryear{Guetta}{2001}]{Guetta2001} Guetta, D., Spada, M. \& Waxman, E. 2001, \apj, 557, 399
\bibitem[\protect\citeauthoryear{Hascoet}{2014}]{Hascoet2014} Hasco{\"e}t, R., Beloborodov, A. M., Daigne, F., et al. 2014, \apj, 782, 5
\bibitem[\protect\citeauthoryear{Hascoet}{2015}]{Hascoet2015} Hasco{\"e}t, R., Vurm, I. \& Beloborodov, A. M.\ 2015, \apj, 813, 63
\bibitem[\protect\citeauthoryear{Jin}{2007}]{Jin2007} Jin, Z. P. \& Fan, Y. Z. 2007, MNRAS, 378, 1043
\bibitem[\protect\citeauthoryear{Jin}{2010}]{Jin2010} Jin, Z. P., Fan, Y. Z. \& Wei, D. M. 2010, \apj, 724, 861
\bibitem[\protect\citeauthoryear{Kann}{2006}]{Kann2006} Kann, D. A., Klose, S., \& Zeh, A. 2006, \apj, 641, 993
\bibitem[\protect\citeauthoryear{Kann}{2010}]{Kann2010} Kann, D. A., Klose, S., Zhang, B., et al. 2010, \apj, 720, 1513
\bibitem[\protect\citeauthoryear{Kobayashi}{2000}]{Kobayashi2000} Kobayashi, S. 2000, \apj, 545, 807
\bibitem[\protect\citeauthoryear{Kobayashi}{2000}]{Kobayashi&Sari2000} Kobayashi, S. \& Sari, R. 2000, \apj, 542, 819
\bibitem[\protect\citeauthoryear{Kobayashi}{2001}]{Kobayashi&Sari2001} Kobayashi, S. \& Sari, R. 2001, \apj, 551, 934
\bibitem[\protect\citeauthoryear{Kobayashi}{2003}]{Kobayashi2003} Kobayashi, S. \& Zhang, B. 2003, \apj, 597, 455
\bibitem[\protect\citeauthoryear{Kumar}{2000}]{Kumar2000} Kumar, P. 2000, \apj, 538, L125
\bibitem[\protect\citeauthoryear{Kumar}{2015}]{Kumar2015} Kumar P. \& Zhang B. 2015, PhR, 561, 1
\bibitem[\protect\citeauthoryear{Laskar}{2009}]{Laskar2009} Laskar, T., Berger, E., Zauderer, B. A., et al. 2013, \apj, 776, 119
\bibitem[\protect\citeauthoryear{Lei}{2011}]{Lei2011} Lei, H. D. Wang, J. Z., L\"u, J. \& Zou, Y. C. 2011, ChPhL, 28, 129801
\bibitem[\protect\citeauthoryear{Lei}{2009}]{Lei2009} Lei, W. H., Wang, D. X., Zhang, L., et al. 2009, \apj, 700, 1970
\bibitem[\protect\citeauthoryear{Lei}{2011}]{Lei2013} Lei, W. H., Zhang, B. \& Liang, E. W. 2013, \apj, 765, 125
\bibitem[\protect\citeauthoryear{Li}{2012}]{Li2012} Li, L., Liang, E. W., Tang, Q. W., et al. 2012, \apj, 758, 27
\bibitem[\protect\citeauthoryear{Li}{2010}]{Li2010} Li, Z. 2010, \apj, 709, 525
\bibitem[\protect\citeauthoryear{Liang}{2013}]{Liang2013} Liang, E., Li, L., Gao, H., et al. 2013, \apj, 774, 13
\bibitem[\protect\citeauthoryear{Liang}{2006}]{Liang2006} Liang, E. \& Zhang, B. 2006, \apj, 638, L67
\bibitem[\protect\citeauthoryear{Liang}{2010}]{Liang2010} Liang, E. W., Yi, S. X., Zhang, J., et al. 2010, \apj, 725, 2209
\bibitem[\protect\citeauthoryear{Lithwick}{2001}]{Lithwick2001} Lithwick, Y. \& Sari, R. 2001, \apj, 555, 540
\bibitem[\protect\citeauthoryear{Lloyd}{2004}]{Lloyd2004} Lloyd-Ronning, N. M. \& Zhang, B. 2004, \apj, 613, 477
\bibitem[\protect\citeauthoryear{Lv}{2012}]{Lv2012} L\"u, J., Zou, Y. C., Lei, W. H., et al. 2012, \apj, 751, 49
\bibitem[\protect\citeauthoryear{Meszaros}{1997}]{Meszaros1997} M{\'e}sz{\'a}ros, P. \& Rees M. J. 1997, \apj, 476, 232
\bibitem[\protect\citeauthoryear{Meszaros}{1999}]{Meszaros1999} M{\'e}sz{\'a}ros, P. \& Rees, M. J. 1999, MNRAS, 306, L39
\bibitem[\protect\citeauthoryear{Meszaros}{1998}]{Meszaros1998} M{\'e}sz{\'a}ros, P., Rees, M. J. \& Wijers, R. A. M. J. 1998, \apj, 499, 301
\bibitem[\protect\citeauthoryear{Metzger}{2011}]{Metzger11} Metzger, B. D., Giannios, D., Thompson, T. A., et al. 2011, MNRAS, 413, 2031
\bibitem[\protect\citeauthoryear{Mimica}{2010}]{Mimica10} Mimica, P., Giannios, D. \& Aloy, M. A. 2010, MNRAS 407, 2501
\bibitem[\protect\citeauthoryear{Molinari}{2007}]{Molinari2007} Molinari, E., Vergani, S. D., Malesani, D., et al. 2007, A\&A, 469, L13
\bibitem[\protect\citeauthoryear{Mu}{2016}]{Mu2016} Mu, H. J., Lin, D. B., Xi, S. Q., et al. 2016, \apj, 831, 111
\bibitem[\protect\citeauthoryear{Nagataki}{2009}]{Nagataki2009} Nagataki, S. 2009, \apj, 704, 937
\bibitem[\protect\citeauthoryear{Nagataki}{2011}]{Nagataki2011} Nagataki, S. 2011, PASJ, 63, 1243
\bibitem[\protect\citeauthoryear{Narayan}{1992}]{Narayan1992} Narayan, R., Paczynski, B. \& Piran, T. 1992, \apj, 395, L83
\bibitem[\protect\citeauthoryear{Nardini}{2006}]{Nardini2006} Nardini, M., Ghisellini, G., Ghirlanda, G., et al. 2006, A\&A, 451, 821
\bibitem[\protect\citeauthoryear{Oates}{2009}]{Oates2009} Oates, S. R., Page, M. J., Schady, P., et al. 2009, MNRAS, 395, 490
\bibitem[\protect\citeauthoryear{Panaitescu}{2000}]{Panaitescu2000} Panaitescu, A. \& Kumar, P. 2000, \apj, 543, 66
\bibitem[\protect\citeauthoryear{Panaitescu}{2001}]{Panaitescu2001} Panaitescu, A. \& Kumar, P. 2001, \apj, 560, L49
\bibitem[\protect\citeauthoryear{Panaitescu}{2001}]{Panaitescu2002} Panaitescu, A. \& Kumar, P. 2002, \apj, 571, 779
\bibitem[\protect\citeauthoryear{Panaitescu}{2004}]{Panaitescu2004} Panaitescu, A. \& Kumar, P. 2004, MNRAS, 353, 511
\bibitem[\protect\citeauthoryear{Peng}{2014}]{Peng2014} Peng, F. K., Liang, E. W., Wang, X. Y., et al. 2014, \apj, 795, 155
\bibitem[\protect\citeauthoryear{Perley}{2008}]{Perley2008} Perley, D. A., Bloom, J. S., Butler, N. R., et al. 2008, \apj, 672, 449
\bibitem[\protect\citeauthoryear{Perley}{2010}]{Perley2010} Perley, D. A., Bloom, J. S., Klein, C. R., et al. 2010, MNRAS, 406, 2473
\bibitem[\protect\citeauthoryear{Perley}{2014}]{Perley2014} Perley, D. A., Cenko, S. B., Corsi, A., et al. 2014, \apj, 781, 37
\bibitem[\protect\citeauthoryear{Perley}{2013}]{Perley2013} Perley, D. A., Levan, A. J., Tanvir, N. R., et al. 2013, 778, 128
\bibitem[\protect\citeauthoryear{Popham}{1999}]{Popham1999} Popham, R., Woosley, S. E. \& Fryer, C. 1999, \apj, 518, 356
\bibitem[\protect\citeauthoryear{Rykoff}{2009}]{Rykoff2009} Rykoff, E. S., Aharonian, F., Akerlof, C. W., et al. 2009, \apj, 702, 489
\bibitem[\protect\citeauthoryear{Sari}{1996}]{Sari1996} Sari, R., Narayan, R. \& Piran, T. 1996, \apj, 473, 204
\bibitem[\protect\citeauthoryear{Sari}{1995}]{Sari1995} Sari, R. \& Piran, T. 1995, \apj, 455, L143
\bibitem[\protect\citeauthoryear{Sari}{1999}]{Sari1999a} Sari, R. \& Piran, T. 1999a, \apjl, 517, L109
\bibitem[\protect\citeauthoryear{Sari}{1999})]{Sari1999b} Sari, R. \& Piran, T. 1999b, \apjl, 520, 641
\bibitem[\protect\citeauthoryear{Sari}{1998}]{Sari1998} Sari, R., Piran, T. \& Narayan, R. 1998, \apj, 497, L17
\bibitem[\protect\citeauthoryear{Schlafly}{2011}]{Schlafly2011} Schlafly, E. F. \& Finkbeiner, D. P. 2011, \apj, 737, 103
\bibitem[\protect\citeauthoryear{Soderberg}{2002}]{Soderberg2002} Soderberg, A. M. \& Ramirez-Ruiz, E. 2002, MNRAS, 330, L24
\bibitem[\protect\citeauthoryear{Soderberg}{2003}]{Soderberg2003} Soderberg, A. M. \& Ramirez-Ruiz, E. 2003, MNRAS, 345, 854
\bibitem[\protect\citeauthoryear{Sonbas}{2015}]{Sonbas2015} Sonbas, E., MacLachlan, G. A., Dhuga, K. S., et al. 2015, \apj, 805, 86
\bibitem[\protect\citeauthoryear{Tang}{2015}]{Tang2015} Tang, Q. W., Peng, F. K., Wang, X. Y. \& Tam, P. H. T. 2015, \apj, 806, 194
\bibitem[\protect\citeauthoryear{Thompson}{1994}]{Thompson1994} Thompson, C. 1994, MNRAS, 270, 480
\bibitem[\protect\citeauthoryear{Uhm}{2012}]{Uhm2012} Uhm, Z. L, Zhang, B., Hasco\"{e}t, R, et al. 2012, \apj, 761, 147
\bibitem[\protect\citeauthoryear{Usov}{1992}]{Usov1992} Usov, V. V. 1992, Nature, 357, 472
\bibitem[\protect\citeauthoryear{van}{2014}]{van2014} van Eerten, H. J.\ 2014, \mnras, 445, 2414
\bibitem[\protect\citeauthoryear{Vestrand}{2014}]{Vestrand2014} Vestrand, W. T., Wren, J. A., Panaitescu, A., et al. 2014, Science, 343, 38
\bibitem[\protect\citeauthoryear{Wang}{2002}]{Wang2002} Wang, D. X., Xiao, K. \& Lei, W. H. 2002, MNRAS, 335, 655
\bibitem[\protect\citeauthoryear{Wang}{2013}]{Wang2013} Wang, X., Liang, E., Li, L., et al. 2013, \apj, 774, 132
\bibitem[\protect\citeauthoryear{Wang}{2015}]{Wang2015} Wang, X., Zhang, B., Liang, E., et al. 2015, \apjs, 219, 9
\bibitem[\protect\citeauthoryear{Woods}{1995}]{Woods1995} Woods, E. \& Loeb A. 1995, \apj, 453, 583
\bibitem[\protect\citeauthoryear{Woosley}{1993}]{Woosley1993} Woosley, S. E. 1993, \apj, 405, 273
\bibitem[\protect\citeauthoryear{Wu}{2003}]{Wu2003} Wu, X. F., Dai, Z. G., Huang, Y. F. \& Lu, T. 2003, MNRAS, 342, 1131
\bibitem[\protect\citeauthoryear{Wu}{2004}]{Wu2004} Wu, X. F., Dai, Z. G., Huang, Y. F. \& Ma, H. T. 2004, Chinese J. Astron. Astrophys., 4, 455
\bibitem[\protect\citeauthoryear{Xue}{2009}]{Xue2009} Xue, R. R., Fan Y. Z., \& Wei D. M., 2009, A\&A, 498, 671
\bibitem[\protect\citeauthoryear{Yi}{2017}]{Yi2017} Yi, S. X., Lei, W. H., Zhang, B., et al. 2017, JHEAp, 13, 1
\bibitem[\protect\citeauthoryear{Yi}{2013}]{Yi2013} Yi, S. X., Wu, X. F. \& Dai. Z. G. 2013, \apj, 776, 120
\bibitem[\protect\citeauthoryear{Yi}{2015}]{Yi2015} Yi, S. X., Wu, X. F., Wang, F. Y. \& Dai. Z. G., 2015, \apj, 807, 92
\bibitem[\protect\citeauthoryear{Zaninoni}{2013}]{Zaninoni2013} Zaninoni, E., Grazia, B. M., Margutti, R., Oates, S. \& Chincarini, G. 2013, A\&A, 557, 12
\bibitem[\protect\citeauthoryear{Zhang}{2003}]{Zhang2003} Zhang, B., Kobayashi, S. \& M{\'e}sz{\'a}ros, P.\ 2003, \apj, 595, 950
\bibitem[\protect\citeauthoryear{Zhang}{2001}]{Zhang2001} Zhang, B. \& M{\'e}sz{\'a}ros, P. 2001, \apj, 552, L35
\bibitem[\protect\citeauthoryear{Zhang}{2002}]{Zhang2002} Zhang, B. \& M{\'e}sz{\'a}ros, P. 2002, \apj, 581, 1236
\bibitem[\protect\citeauthoryear{Zhao}{2011}]{Zhao2011} Zhao, X. H., Li, Z. \& Bai, J. M. 2011, \apj, 726, 89
\bibitem[\protect\citeauthoryear{Zou}{2011}]{Zou2011} Zou, Y. C., Fan, Y. Z. \& Piran, T. 2011, \apjl, 726, L2
\bibitem[\protect\citeauthoryear{Zou}{2010}]{Zou2010} Zou, Y. C. \& Piran, T. 2010, MNRAS, 402, 1854
\bibitem[\protect\citeauthoryear{Zou}{2005}]{Zou2005} Zou, Y. C., Wu, X. F. \& Dai, Z. G. 2005, MNRAS, 363, 93
\end{thebibliography}




%
%
%

\bsp	
\label{lastpage}
\end{document}